\documentclass[twocolumn,aps,superscriptaddress]{revtex4-1}
\usepackage{graphicx}
\usepackage{amsmath}
\usepackage{amssymb}

\def\ben{\begin{equation}}
\def\een{\end{equation}}
\def\sss{\scriptscriptstyle\rm}

\def\c{_{\sss C}}
\def\s{_{\sss S}}
\def\xc{_{\sss XC}}

\def\hxc{_{\sss HXC}}

\def\tot{_{\sss tot}}
\def\up{_\uparrow}
\def\dn{_\downarrow}
\def\tk{$T_{\rm K}$}
\def\o{\omega}
\def\e{\varepsilon}
\def\g{\Gamma}
\def\drop{\delta V\s/\delta V}

\def\br{{\bf r}}

\begin{document}
\title{Density functional description of Coulomb blockade:\\  Adiabatic or
dynamic exchange-correlation?} 
\author{Zhen-Fei Liu}
\affiliation{Molecular Foundry and Materials Sciences Division, Lawrence 
Berkeley National Laboratory, Berkeley, California 94720, USA}
\author{Kieron Burke}
\affiliation{Departments of Chemistry and Physics, University of California, 
Irvine, California 92697, USA}
\date{\today}

\begin{abstract}
Above the Kondo temperature, the Kohn-Sham zero-bias conductance of an Anderson
junction  {has been shown to} completely miss the Coulomb blockade.
Within a standard model for the spectral function, we deduce
a parameterization for both the onsite exchange-correlation potential and the
bias drop as a function of the site occupation  {that applies
for all correlation strengths.
We use our results to sow doubt on the common interpretation of such corrections
as arising from dynamical exchange-correlation contributions.}
\end{abstract}

\maketitle

\section{Introduction}

Electron transport \cite{HJ07} through molecular junctions is usually 
formulated using non-equilibrium Green's function (NEGF)\cite{MW92}
or an equivalent scattering formalism \cite{FL81}. 
In atomistic simulations of electron transport \cite{TGW01,transiesta},
density functional theory (DFT) \cite{KS65,B12} is often
used in combination with NEGF, producing the 
standard model\cite{KCBC08} in which
the interacting one-body Green's function and coupling
matrices are replaced by their Kohn-Sham (KS) counterparts
\cite{TGW01,transiesta}. 
Whether or not ground-state KS-DFT can in principle, under certain
conditions,  produce the exact conductance
remains open\cite{KCBC08},
even at zero temperature and in the linear response regime. 
Thus there are two sources of error which can be difficult
to disentangle: (i) replacing the interacting
one-body Green's function with the KS Green's function, and
(ii) replacing the exact KS Green's function by one
from an approximate functional\cite{TFSB05}. 

Model Hamiltonians provide excellent testbeds for this purpose,
especially when highly accurate results are known and
can serve as benchmarks for larger KS-DFT calculations. 
The Anderson impurity model \cite{A61} has attracted much attention
recently\cite{SK11, BLBS12, LBBS12, TSE12, KS13, SK13, ES13, ES13b}.
Due to its generality, broad applicability, and
exact solvability, studying transport through an Anderson junction 
is useful in understanding errors in atomistic
calculations and in constructing accurate functionals for electron transport.

The KS system is defined as a junction with $U=0$ (see below), but whose
onsite energy is chosen to make its site occupation match that of the interacting
junction, which can be found exactly using the Bethe ansatz\cite{WT83}.
At zero temperature and in the linear response regime, 
the KS conductance, i.e., the conductance of non-interacting electrons
in the KS potential, is exact\cite{BLBS12,SK11,TSE12}, i.e., the first
error is zero.  
This can be attributed to the Friedel-Langreth sum rule \cite{F58,L66},
which implies the transmission is a simple function of occupation.
By reverse engineering of the exact solution\cite{BLBS12}, one can
study which features must be present in any approximation
in order to generate an accurate transmission\cite{LBBS12}. For example, the
derivative discontinuity\cite{PPLB82} is crucial as correlations grow,
and how rapidly it is approached as $U$ grows is determined by the charge
susceptibility of the system at particle-hole symmetry \cite{LBBS12}.
 {This derivative discontinuity has also been shown to be 
related to more general blockade phenomena \cite{CBKR07,KSKV10}.}

However, all this changes above the Kondo temperature\cite{PG04}, when the
Kondo transmission peak is negligible, and the sum rule no longer applies.
Accurate solutions are available \cite{HJ07}, and the spectral function 
exhibits two Coulomb
peaks in the strongly correlated regime. Between the two peaks, the
linear-response conductance is very low due to Coulomb blockade. 
This temperature regime better mimics what happens
in real molecular junctions than the zero-temperature limit where a Kondo plateau dominates. But 
the  {low-temperature} KS conductance \emph{always} satisfies the 
sum rule even when the physical conductance does not, and so
is qualitatively incorrect between the two Coulomb peaks. 
Having spent years figuring out how to get the Kondo plateau into DFT,
the difficulty is now to get rid of it at finite temperature {\cite{note2}.}

It has long been argued\cite{SZVD05,KBE06,VD09} that time-dependent DFT 
produces \emph{dynamic} corrections to the
KS conductance, and that these are {\em necessary} to produce an accurate transmission.
Making this identification, Ref. \cite{KS13} shows that
a simple model for the exchange-correlation (XC) kernel of 
time-dependent density functional theory (TDDFT) 
reproduces the right corrections,
at least in the strongly correlated limit. This was further shown to
depend, somewhat surprisingly, on the onsite occupation alone.

 {Here, we argue that the corrections that turn the KS conductance into 
the true conductance of an Anderson junction can plausibly be
considered as a {\em  non-local static} correction that in principle could be
extracted from ground-state DFT.   This possibility was suggested long ago \cite{KBE06}
as the only alternative to dynamic TDDFT effects for altering a resonance
in the transmission of such a junction.   The origin of such an effect, within
ground-state DFT, is the well-documented counteracting XC field that
is significant for certain
molecules and solids in response to a long-range electric field.  The origin
of this field is the localization of orbitals on specific sites and the appearance
of step-like features in the induced exchange-correlation potential that (correctly) reduce the
polarization, relative to that of standard local, semilocal, and hybrid functionals\cite{KKP04}.
These steps are reasonably accurately given by Hartree-Fock or 
optimized effective potential (OEP) calculations,
because of their explicit orbital dependence\cite{KK08}.  They are in fact a field-induced
derivative discontinuity.}

 {To make this argument, we first accurately parametrize the onsite XC potential
for temperatures that are low, but at which there is no Kondo plateau in the 
conductance, i.e., in the Coulomb blockade regime.   
Our parametrization is designed to work for all correlation
strengths, not just for strong correlation.  Next, we accurately parametrize the
XC bias drop that ensures that a static KS calculation reproduces the physical
conductance, again including both weak and strong correlation regimes.
Finally, we explain how this can be interpreted in terms of the known counteracting
XC fields.}

\section{Anderson Junction at Low Temperature Limit}

To begin, the Hamiltonian for the Anderson junction\cite{A61}
consists of an interacting impurity
site coupled to identical featureless left and right leads:
\ben
{\cal H}=\e\, \hat{n}+U\, \hat{n}\up\hat{n}\dn + {\cal H}_{\rm leads} + {\cal H}_{\rm T}
\een
with $\e$ the on-site energy and $U$ the Coulomb repulsion when the site is
doubly occupied, $\hat{n}=\hat{n}\up+\hat{n}\dn$.
Here ${\cal H}_{\rm leads}$ is
the Hamiltonian of the two leads, and ${\cal H}_{\rm T}$ is their
coupling to the site.  
In the wide band limit, the effect of tunneling is incorporated into an
energy-independent constant $\Gamma$ \cite{note1}. 
There are two dimensionless parameters: $(\mu-\e)/U$, with $\mu$ the
chemical potential of the leads, which moves the system on- and off-resonance,
and $u=U/\Gamma$, which switches the system between weakly and
strongly correlated.  Below the Kondo temperature \tk, a characteristic 
temperature
dependent on $u$, the exact spectral function has two Coulomb peaks
and one Kondo peak whose width is related to \tk \cite{SGSJ90}, and the sum 
rule applies. 
Above \tk, the Kondo peak disappears, the sum rule is violated, and the
conductance comes from the two Coulomb peaks only \cite{SGSJ90}. 

Although no exact solution is available above \tk,
the Green's function on the central impurity site can be accurately approximated
as \cite{HJ07} 
\ben
G(\omega)=\sum_{i=1,2}\frac{n_i}{\omega-\e_i+i\Gamma/2},
\label{gom}
\een
where $\e_1=\e, \e_2=\e+U, n_1=1-n/2,$ and $n_2=n/2$, with
$n\equiv\left<\hat{n}\c\right>$ the occupation on the impurity site, and
the coupling to the leads causes the broadening of $\Gamma/2$.
Throughout this work, we use Eq. \eqref{gom} as an \emph{ansatz} for the exact 
solution, as it captures the right physics and accurately mimics the numerical
exact solution above \tk \cite{HJ07}. We also introduce a specific low 
temperature limit, namely $\beta^{-1} \to 0$, but $\beta^{-1} \gg k_BT_{\rm K}$,
where $\beta$ is the inverse temperature and $k_B$ the Boltzmann constant.
Because $T_{\rm K}$ depends exponentially on the parameters,
it is typically much smaller than any other temperature scale.
We are in
a regime above \tk, but at temperatures far smaller than
any of the other energy scales of the problem.
For example, Kurth and Stefanucci
\cite{KS13} discussed a case where $U=10,\Gamma=1,$ and temperature $\tau=0.1$ 
($T_{\rm K}=0.06$). We have checked that by taking our low temperature limit,
one obtains essentially the same numerical results as using $\tau=0.1$ in the Fermi 
function. In the rest of this work, we consider only this limit, which greatly 
simplifies the derivation and analytical forms are then available for the 
quantities of interest.  {We note that we use the same Eq. \eqref{gom}
as Ref. \cite{KS13} for the benchmark solution above \tk, to facilitate comparison of
our approach (static correction) and that in Ref. \cite{KS13} (dynamic correction).
Ref. \cite{KS13} studied these effects for a range of temperatures but only for
strong correlation; here we are interested in {\em all} correlation strengths, and
find that low temperatures are well approximated by a specific low-temperature
limit.  In this limit, we find accurate analytic parametrizations with relative
ease.}

The spectral function $A(\omega)=-2\mbox{Im}G(\omega)$ depends on the 
occupation $n$ on the impurity site. Therefore $n$ must be determined self-consistently:
\ben
n=
\frac{1}{\pi}\int_{-\infty}^{\mu}\,d\omega A(\omega)
=\frac{\pi-2\theta_1}{\pi+\theta_2-\theta_1},
\label{den}
\een
where $\tan\theta_i=2(\e_i-\mu)/\Gamma$ and,
in the low temperature limit, levels are fully occupied up to
$\mu$.  
Thanks to this limit,  Eq. (\ref{den})
is a closed-form for $n$ in terms of the
parameters.
The linear-response transmission is then \cite{KS13}:
\ben
T=-\frac{\Gamma}{2}
\int_{-\infty}^{\infty}\frac{d\omega}{2\pi}\frac{df}{d\omega}\, A(\omega) \xrightarrow{\beta^{-1}\to 0} \frac{\Gamma}{4\pi}A(\omega=\mu).
\label{ext}
\een

\begin{figure}[htb]
\begin{center}
\includegraphics[width=3.5in]{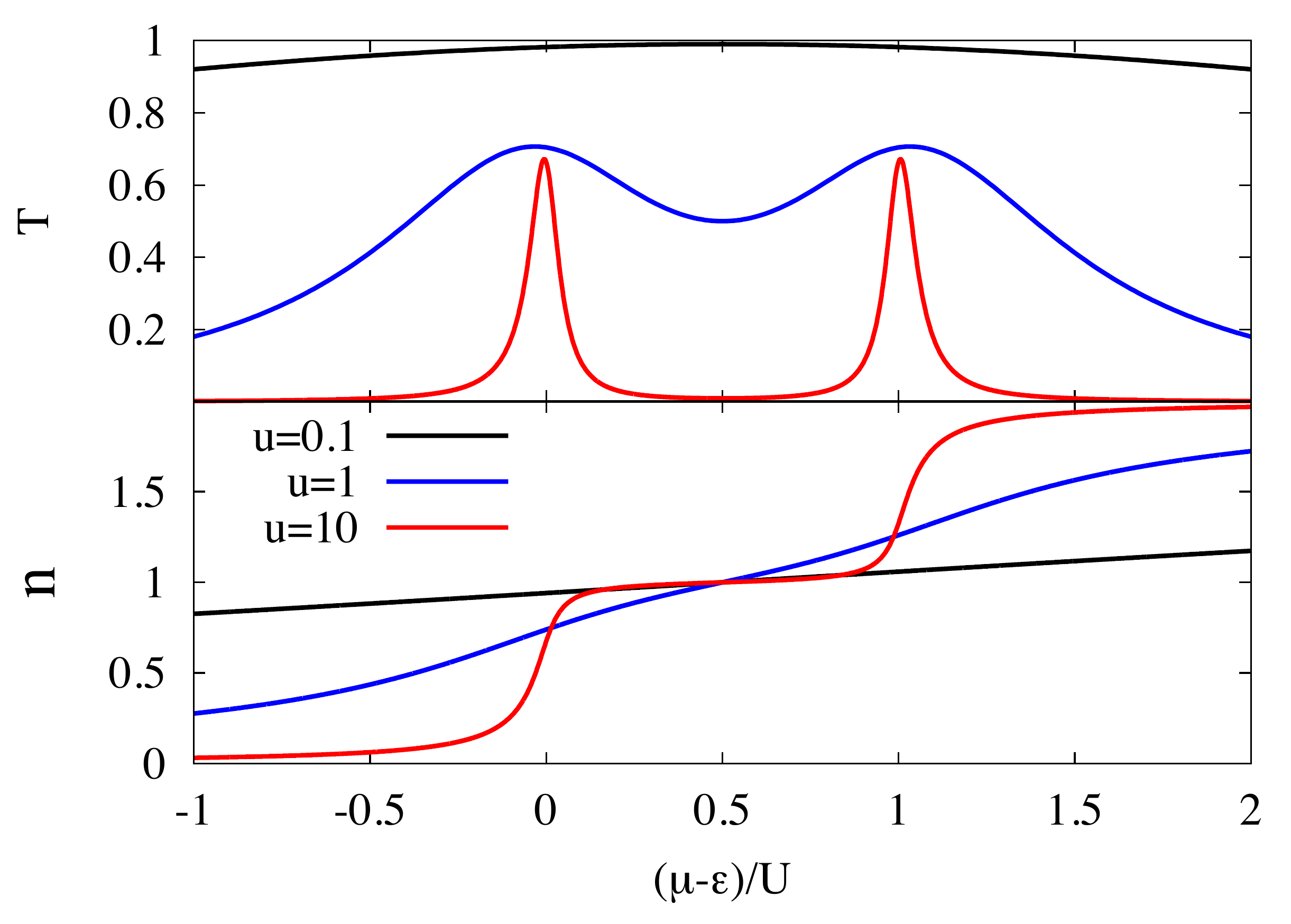}
\caption{Upper panel: Conductance $T$ as a function of $(\mu-\e)/U$, plotted in unit of 
$G_0=2e^2/h=1/\pi$, the conductance quantum. 
lower panel: $n$ as a function of $(\mu-\e)/U$. Different $u=U/\g$ are used.}
\label{fig1}
\end{center}
\end{figure}

In Fig. \ref{fig1}, $n$ and $T$
are plotted as a function of $\mu$, for several values of $u$.
For $U \gg \g$, a plateau in $n$ develops, i.e., a Coulomb blockade,
which corresponds to the low conductance region between the two Coulomb peaks
at $\mu \approx \epsilon$ and $\mu \approx \epsilon + U$.
This is very different from the Kondo regime, where there is always
a Kondo plateau in conductance, 
as discussed in Ref. \cite{BLBS12}.

\section{Kohn-Sham Anderson Junction}

We now construct a KS Anderson junction, i.e., one with $U=0$,
that generates the same occupation $n$ as the interacting system by replacing $\e$ with 
$\e\s[n]=\e+v\hxc[n]$, and analyze its conductance. Because the occupation on the 
impurity site is a number, the functional dependence is simply a function of $n$.
The KS Green's function has only one pole:
\ben
G\s(\omega)=\frac{1}{\omega-\e\s[n]+i\Gamma/2}.
\label{ksgreen}
\een
The Hartree-exchange-correlation (HXC) potential $v\hxc[n]$ is defined such that:
\ben
n=\frac{1}{\pi}\int_{-\infty}^{\mu}\,d\omega A\s(\omega),
\label{den2}
\een
where $A\s(\omega)=-2\mbox{Im}G\s(\omega)$ is the spectral function of the 
KS system. By definition\cite{CFSB15}, the
KS occupation matches the physical one, i.e., 
the left hand sides of Eqs. \eqref{den} and \eqref{den2}
are identical, for a given set of $\mu,\e,U,$ and $\Gamma$. Applying this condition
yields:
\ben
v\hxc=\mu-\e-\frac{\g}{2}\tan\left[\frac{\pi}{2}(n-1)\right].
\label{vhxc}
\een
Note that this is {\em insufficient} to define $v\hxc[n]$ because of the
presence of $\mu-\e$  [If Eq. (\ref{den}) could be inverted to find $\mu-\e$
as an explicit function of $n$, as is trivial numerically, this would suffice].
In Ref. \cite{KS13}, 
reverse engineering to find $v\hxc[n]$ is done at three different 
temperatures above \tk, but for a fixed  {large} $u$ (=10). Here the 
low temperature limit simplifies the algebra, and we study the $u$-dependence
 {explicitly.}
Following the ideas of Ref. \cite{LBBS12}, we  {parametrize} $v\hxc[n]$ as
\ben
v\hxc^{\rm app}[n]=\frac{U}{2}\left(1+\frac{2}{\pi}\tan^{-1}[\sigma(u)(n-1)]
\right),
\label{vhxcfunc}
\een
where $\sigma(u)$ is a parameter, determined by the exact condition of charge 
susceptibility at particle-hole symmetry\cite{H97}:
\ben
\chi=U \left.\frac{\partial n}{\partial \mu}\right|_{n=1}
=\frac{4u}{(1+u^2)(\pi+2\tan^{-1} u)}, 
\label{chi}
\een
where Eq. \eqref{den} was used. 
The derivative is taken at the 
particle-hole symmetry point, $n=1$, or equivalently where
$\epsilon=\mu-U/2$. Imposing
this condition fixes the parameter $\sigma$ in Eq. \eqref{vhxcfunc}
as $\sigma(u)=\pi/\chi-\pi^2/(4u)$. 
 {This procedure could be simply generalized to finite temperature, by using
the finite-temperature susceptibility, but this is not the main concern
of the present work. We briefly show and discuss finite-temperature effects in
Section \ref{fte} using the numerical solution, without analytically parametrizing
the temperature-dependent functional.}

\begin{figure}[htb]
\begin{center}
\includegraphics[width=3.5in]{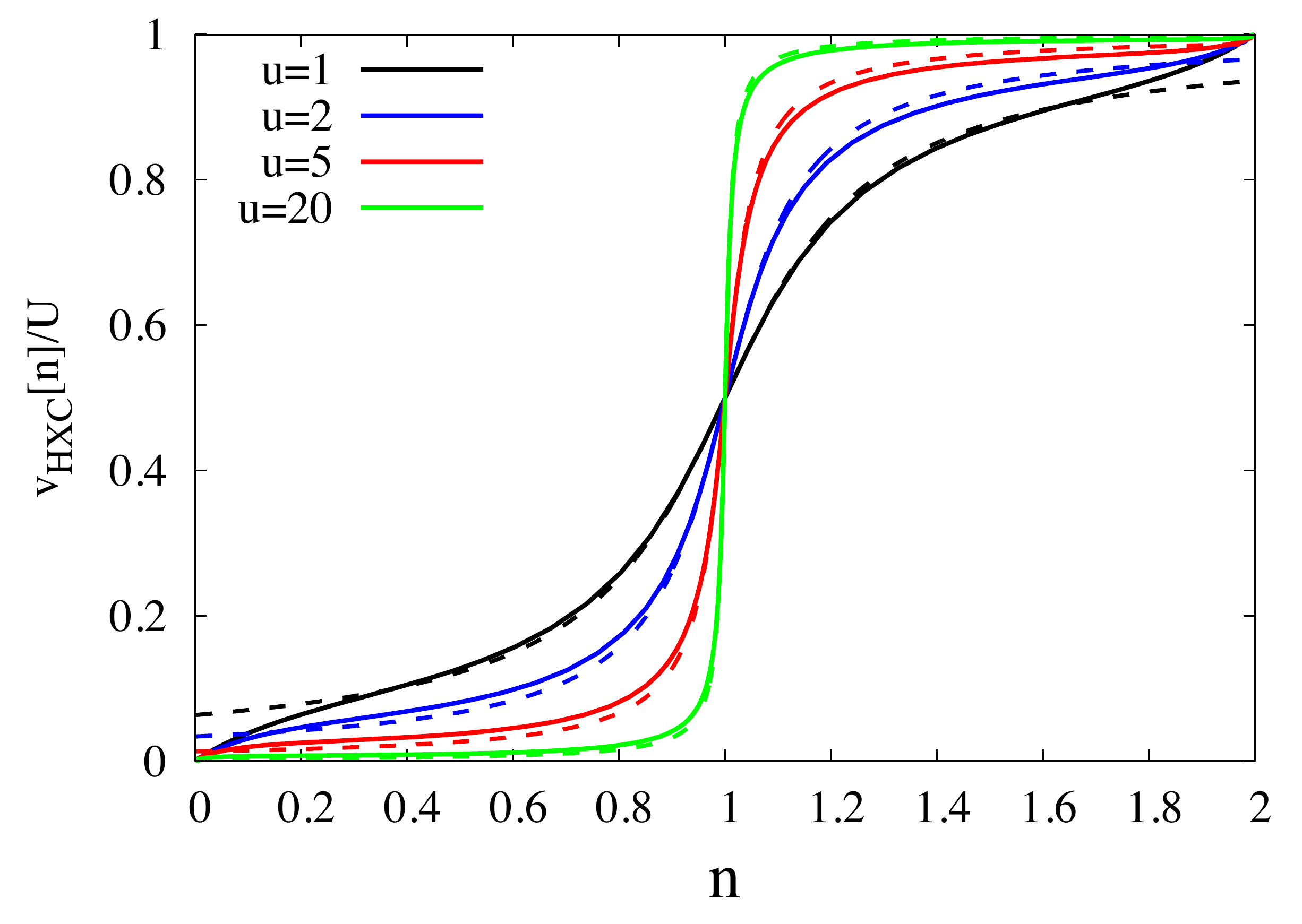}
\caption{$v\hxc[n]$ as a function of $n$ for various $u=U/\Gamma$. Solid lines are
results from Eq. (\ref{vhxc}) with $n$ calculated from Eq. \eqref{den}, dashed lines are the approximation of Eq. \eqref{vhxcfunc}.}
\label{vxc}
\end{center}
\end{figure}
In Fig. \ref{vxc}, both the $v\hxc[n]$ that precisely reproduces the
onsite occupation determined from Eq. \eqref{den} and the  {parametrization of} 
Eq. \eqref{vhxcfunc} are plotted as a 
function of $n$, for different $u$. For $U \gg \g$, a step in $v\hxc$ 
develops, reflecting the onset of the
derivative discontinuity as $u\to\infty$\cite{CFSB15}. 
These kinds of features were discussed in Refs. 
\cite{BLBS12,LBBS12}, at zero temperature. In the low temperature limit 
discussed in this work, these rules still apply. 
 {An important point is that, unlike Ref. \cite{KS13}, our parametrization
applies for all values of $U$ (both weak and strong correlation), and 
appears to fail only when the site is almost entirely empty or doubly-occupied.
Our results match those of Ref. \cite{KS13} when the temperature is low and
the correlation is strong.}

\begin{figure}[htb]
\begin{center}
\includegraphics[width=3.5in]{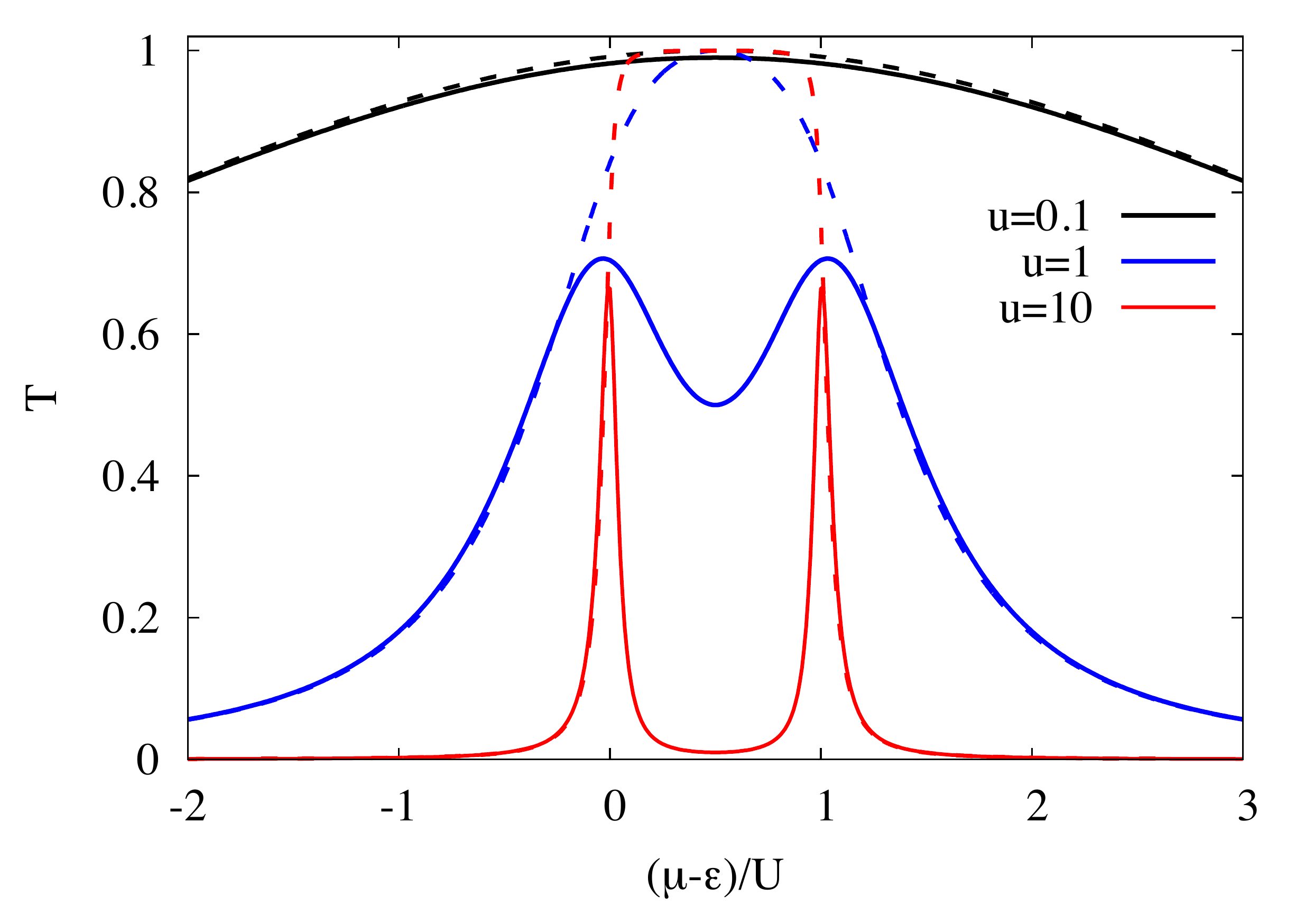}
\caption{Physical conductance (solid lines) from Eq. \eqref{ext} 
and KS conductance (dashed lines) for 
various values of $u=U/\g$,
plotted in unit of $G_0=2e^2/h=1/\pi$, the conductance quantum.}
\label{twot}
\end{center}
\end{figure}
The KS conductance has the same form as in Eq. \eqref{ext}, but with $A$ replaced
by $A\s$. We compare the two quantities in Fig. \ref{twot}. 
Here the KS conductance \emph{always} satisfies the Friedel-Langreth sum rule, 
i.e., $T\s= \sin^2(\pi n/2)$ in the low temperature limit, 
and the plateau is always present, regardless of whether the physical
system is in the Kondo regime (Ref. \cite{BLBS12}) or not (this work). 
However the conductance from Eq. (\ref{ext}) has two 
peaks and is very small between the two peaks for $U \gg \g$
due to the ansatz of Eq. \eqref{gom}. 
For $U \ll \g$, the KS conductance is quite accurate, but for larger $U$,
it is wildly inaccurate.  Whenever correlation is significant (or, equivalently
in this model, the coupling to the leads is weak), the KS conductance
is a large overestimate of the physical conductance.
In particular, it misses entirely the Coulomb blockade effect, producing
no drop at all between the Coulomb peaks.  All this was beautifully
shown in Ref. \cite{KS13}  {for strong correlation for several temperatures.
Although the Friedel-Langreth sum rule does not apply at finite temperatures, 
the discrepancy remains qualitatively the same.}

\section{Static Correction to Kohn-Sham Conductance}

It is no surprise that the KS conductance is not correct under such conditions,
as it is extracted entirely from equilibrium DFT\cite{M65}.
One can formally identify the corrections to the KS conductance\cite{KCBC08},
and capture them in terms of an XC correction to the applied bias:
\ben
\delta I = T\, \delta V =  T\s\, \delta V\s,
\label{curr2}
\een
where $\delta V$ is an applied bias, $\delta V\s=\delta V + 
\delta V\xc$ is the bias experienced by the
KS system, and $\delta I$ is the induced current in both.
The formal result of Ref. \cite{BLBS12} can be stated that $\delta V\xc=0$ due to the 
sum rule at zero temperature
when the Kondo peak is present, and the pioneering work of Ref. \cite{KS13} interpreted
the XC corrections in $\delta V\s$ as dynamical corrections arising
from TDDFT, i.e., corrections that would not appear in any static DFT
calculation.

\begin{figure}[htb]
\begin{center}
\includegraphics[width=3.5in]{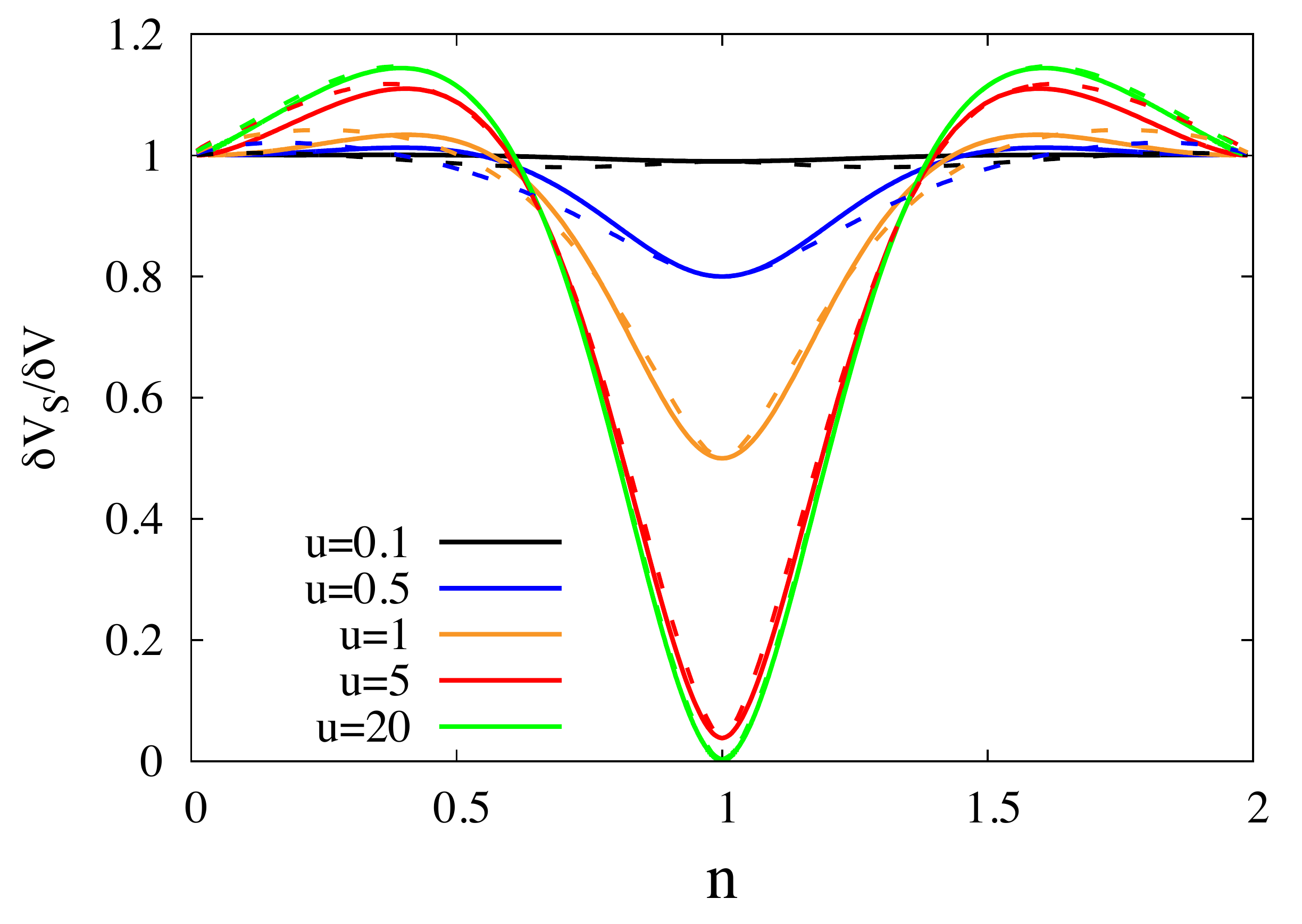}
\caption{$\delta V\s^{\rm app}$ as a function 
of $n$ for various correlation strengths $u=U/\Gamma$.
Solid lines are ``exact'', as defined by Eq. \eqref{dvxcdv}. Dashed lines are approximate, as in Eq. \eqref{dropasn}.}
\label{dvsdv}
\end{center}
\end{figure}

 {Here we suggest that a static correction from ground-state DFT can explain the results 
of Kurth and Stefanucci\cite{KS13} which they interpreted solely in terms of TDDFT.}
We propose the following approximation for a non-local XC bias:
\ben
\frac{\delta V^{\rm app}\s}{\delta V}=
\frac{n_1[1-\cos(2\zeta)]+
{2 n_2}/\left[{1+(2u+\cot\zeta)^2}\right]} {2\sin^2(n_2\pi)}
\label{dropasn}
\een
for $n<1$, where $\zeta=n_2(\pi + 2 n_2 \tan^{-1}u/n_1)$ and 
$n_1$ and $n_2$ are swapped if $n > 1$.
This specific choice will be justified shortly.
For the present, we simply plot it in Fig. \ref{dvsdv}.  For weak correlation at
any filling, or for $n_1$ or $n_2 < 0.5$, this KS bias
is close to the applied bias.
But for $n$ near $1$ and $U \gg \Gamma$, $\delta V\s$ is noticeably smaller than 
$\delta V$.

\begin{figure}[htb]
\begin{center}
\includegraphics[width=3.5in]{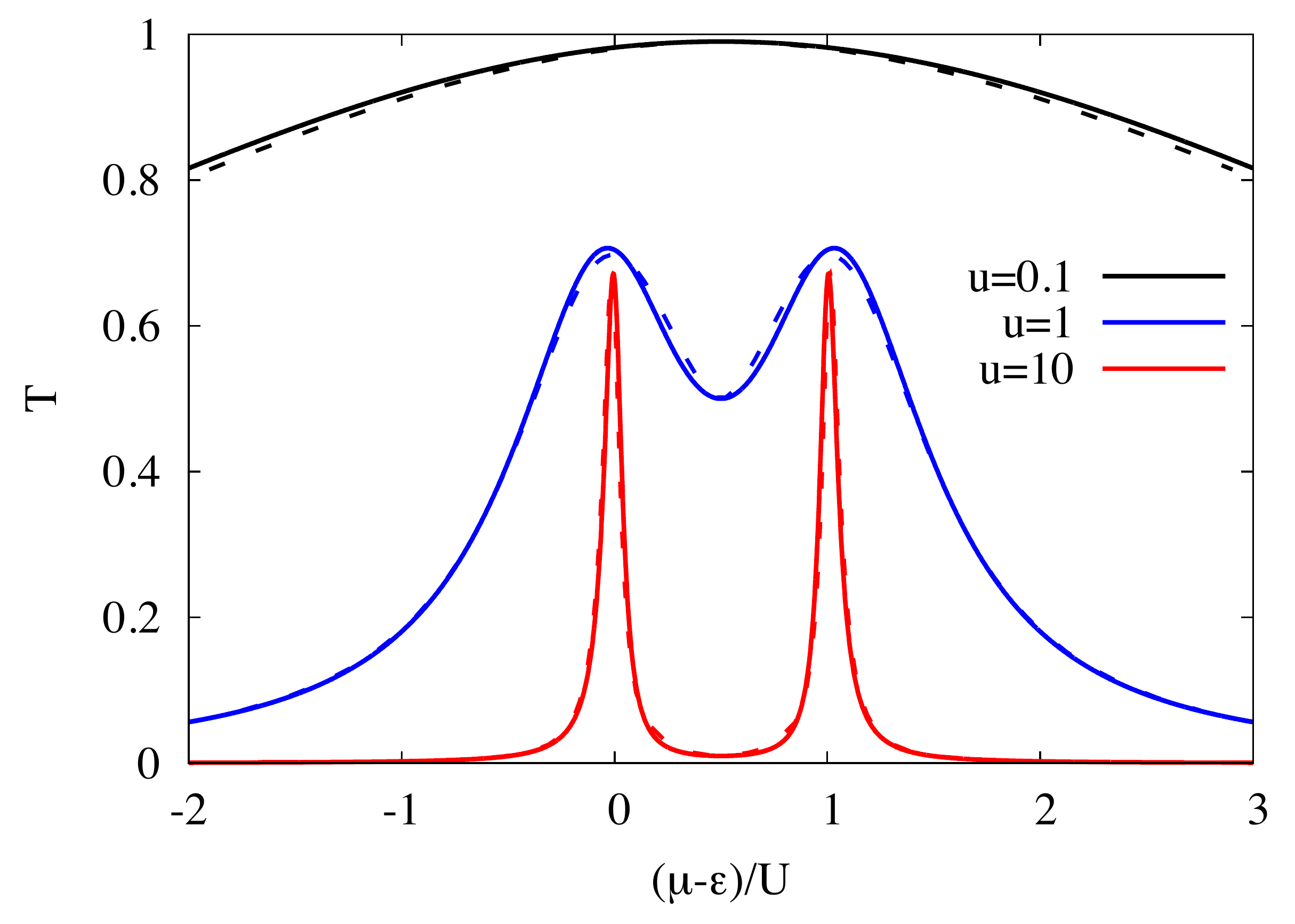}
\caption{Solid lines: physical conductances from Eq. \eqref{ext}. 
Dashed lines: conductances using approximations for both the local HXC potential 
Eq. \eqref{vhxcfunc} and non-local XC bias Eq. \eqref{dropasn}.}
\label{cond}
\end{center}
\end{figure}

Using $\delta V^{\rm app}\s$ in Eq. \eqref{dropasn} with $n$ calculated self-consistently from Eq. \eqref{vhxcfunc}, 
we now plot the approximate conductance in Fig. \ref{cond}.  We see that
it gives us essentially the correct conductance of the Anderson junction, including
the severe corrections needed to reproduce the Coulomb blockade.

To understand how this was achieved, i.e., to derive $\delta V^{\rm app}\s$,
we first ask what
XC drop would {\em exactly} reproduce the conductance implied by
Eq. (\ref{ext}).  From Eq. \eqref{curr2}, we see
\ben
\delta V\s\, A\s(\o=\mu) = \delta V\, A(\o=\mu).
\label{dvxcdv}
\een
The $\drop$ is an implicit functional of $n$ via $(\mu-\e)/U$. 
For example, at particle-hole symmetry, we find explicitly
\ben
\left.\frac{\delta V\s}{\delta V}\right|_{n=1}=\frac{1}{1+u^2},
\hspace{0.2in} \beta^{-1} \to 0.
\label{drop1}
\een
This ``exact'' $\delta V\s$ is plotted in Fig. \ref{dvsdv} as a function of $n$.
The KS bias is close to the applied bias when the site is nearly 
empty or doubly-occupied.  But near $n=1$, as $U$ increases, 
$\drop$ drops rapidly from $1$ to $0$, i.e., the applied bias
is almost entirely screened by the Coulomb blockade.

To use $\delta V\xc$ to correct the KS 
conductance, one needs to express it as an 
explicit functional of the density $n$. The conductance 
is proportional to $A(\omega=\mu)$, which depends
explicitly on $\e$, the external onsite potential (the strength of electron-electron 
interaction, $u=U/\Gamma$, enters into the formula as a parameter). 
We start with Eq. \eqref{den}, and propose the 
following approximation:
\ben
\theta_2(n)\approx (n_1-\frac{1}{2}){\pi}+ 2n_2 \tan^{-1}u,
\label{approx}
\een
for $n < 1$, and $n_1$ and $n_2$ are swapped if otherwise.
Then Eq. \eqref{den} can be inverted to find
$\e[n]=\mu+\g\cot \zeta/2$. After 
substituting $\e[n]$ into Eqs. \eqref{gom} and \eqref{ext}, 
we find our expression for $\drop$.
One can verify that Eq. \eqref{dropasn} satisfies
the relation Eq. \eqref{drop1}.
The approximation used in Ref. \cite{KS13} corresponds to
$\theta_2 \approx \pi/2$ [compared to Eq. \eqref{approx}], which
is accurate only in the large $u$ limit, whereas Eq. (\ref{approx})
works wherever $\delta V\xc$ is significant,  {even for 
small $u$ or weak correlation} [In the language of 
Ref. \cite{KS13}, $R= 2 n_2 \tan^{-1}u/(\pi n_1)$].

\section{Origin of non-local correction}

To understand how the XC bias drop might be extracted from a ground-state
DFT calculation, we review the origin of Eq. \eqref{curr2}.
Start with Kubo response of the system to the external field, and here we follow
the notation in Refs. \cite{KBE06} and \cite{KK01}:
\ben
\delta\mathbf{j}(\mathbf{r};\omega) = \int d\mathbf{r'}\, \hat{\sigma}_{\rm irr}
(\mathbf{r},\mathbf{r'};\omega)\, \delta\mathbf{E}\tot(\mathbf{r'};\omega) 
\label{irrkubo}
\een
where the left-hand side is the current density in response to a perturbing
external electric field at frequency $\omega$,
and $\hat{\sigma}_{\rm irr}$ is the irreducible, frequency-dependent, nonlocal 
conductivity tensor
of the many-electron problem.   Since, in time-dependent
current DFT (TDCDFT), the time-dependent KS system must produce 
the same current-density response, we also have
\ben
\delta\mathbf{j}(\mathbf{r};\omega) 
= \int d\mathbf{r'}\, \hat{\sigma}\s(\mathbf{r},\mathbf{r'};\omega)\,
\left[\delta\mathbf{E}\tot(\mathbf{r'};\omega)+\delta\mathbf{E}\xc(\mathbf{r'};\omega)\right],
\label{kubo}
\een
where
$\delta\mathbf{E}\tot$ contains both the external and Hartree electric fields.
Here $\hat{\sigma}\s$ is the KS conductivity tensor
and $\delta\mathbf{E}\xc$ is the exchange-correlation
contribution to the field felt by the KS system.

The current is defined as an integral of the current density over a cross section:
$\delta I(z;\omega)=\int_S dS \, \delta\mathbf{j}(\mathbf{r};\omega)$,
where $z$ is the direction along
the current flow.  Great care with the order of limits 
must be taken when applying this formula to
our problem\cite{KK01}.
The response formula is true for any $\omega$, but we wish to deduce
the steady-state current.
Thus $\omega$ is kept finite, but reduced to 0 at the
end of the calculation.  In that limit, 
the nonlocal conductivity and the current become coordinate-independent\cite{BS89,KK01,KBE06},
and are just the transmission $T$ in Eq. (10).
Integration of the fields over a large region including the device just yields
the net voltage drop across the device, 
$\delta V = \lim_{L\to\infty} \int_{-L}^L dz \int_S dS \, \delta \mathbf{E}(\br;\omega)$.
Thus we recover Eq. \eqref{curr2}. 

Now comes the tricky part.  For any local (or semilocal) approximation to XC, if the
density deep inside the leads is symmetric far from the molecule, the XC fields must
be also symmetric, 
so that there can be no net XC bias drop.  This is the case in all standard
DFT calculations of transport\cite{KBE06,SAKG06,KCBC08}.  But 
Ref. \cite{KBE06} argues that there are  {\em two} possible
sources of $\delta V\xc$: either highly non-local ground-state effects ($\omega=0$)
or dynamical TDDFT corrections which fail to vanish as $\omega\to 0$
(For any finite system, the dynamical effects must vanish in this limit, but
our system is infinite).

Difficulties for DFT dealing with extended systems in electric fields first arose almost
two decades ago.  The famous GGG papers\cite{GGG95,GGG97} showed that there was, in principle,
a long-range XC counterfield in insulators, that is missed by local and semilocal
approximations.  This is the same contribution that is needed to produce accurate
exciton peaks in the optical response of solids\cite{R94,ORR02}.  Contemporaneously, a similar effect
was found in long chain polymers, whose polarizabilities and hyperpolarizabilities
are greatly overestimated by LDA and GGA calculations\cite{CPGB98,GSGB99,FBLB02,PSB08}.

This counteracting field is easily and accurately approximated by orbital-dependent
functionals, such as Hartree-Fock or exact exchange
 using the OEP formalism\cite{KK08}. It has also been
found that the Vignale-Kohn approximation in TDCDFT, 
even taking the low-frequency
limit, produces a finite correction, although the quantitative accuracy of such
corrections has been questioned in some cases\cite{BBL07,JBG07}.  The work of Ref. \cite{MSB03} 
suggests that either formalism might lead to approximations to the same
feature.  

For DFT calculations of transport through molecular junctions, the TDCDFT correction
has been considered in several calculations and approximations, and 
differences between the KS and true conductance are often attributed to this.
But there is no {\em a priori} reason to discount a non-local static contribution, which
has proved much more straightforward (if expensive) to implement for both the
optical response of insulators and polarizabilities of long-chain polymers.
Thus it is important to understand if the effects seen in Ref. \cite{KS13} can be
understood in this manner.  If so, then they can be searched for in orbital-dependent
calculations of transport, such as GW, etc.

We note here that the crucial element for producing this counteracting XC field
is {\em not} whether or not the system is a metal or insulator (traditional
definitions based on bulk conductance\cite{K64} break down for nanoscale
systems), but rather whether or not charge is localized on a site.  This is
exactly what is being modeled in the Anderson junction.
Furthermore, while TDCDFT is needed to derive Eq. \eqref{curr2}, 
the establishment of the steady current is done by the time-dependence of
the KS equations themselves, without any need for dynamic XC contributions.
As shown by Ref. \cite{KK01}, a steady current is established within even a simple Hartree 
calculation, i.e., one including no XC effects at all.
This shows that the basic feature of a steady current occurs with the time-dependent
KS equations in the absence of any dynamical XC contributions.

\begin{figure}[htb]
\begin{center}
\includegraphics[width=3.5in]{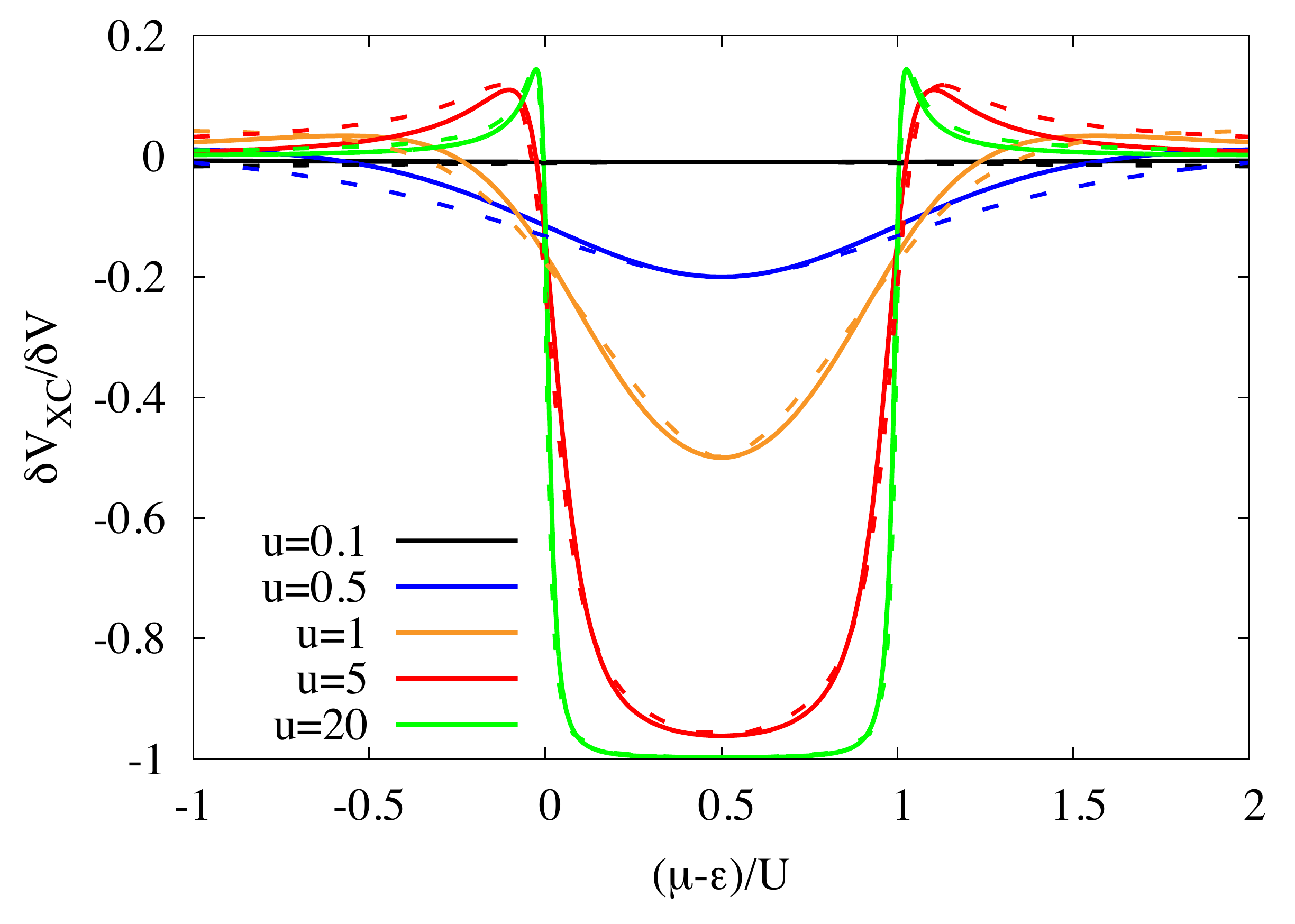}
\caption{$\delta V\xc$ as a function 
of $\mu$ for various correlation strengths $u=U/\Gamma$.
Solid lines are ``exact'' as defined by Eq. \eqref{dvxcdv}
and dashed lines are approximations using both Eqs. \eqref{dropasn} and \eqref{vhxcfunc}.}
\label{figdvxcdv}
\end{center}
\end{figure}

To illustrate how this interpretation makes sense, 
we also plot $\delta V\xc$ in Fig. \ref{figdvxcdv} as a function of $\mu-\e$. 
For large $U$ the XC bias almost completely cancels the applied field,
just as happens for well-separated molecules in response to applied fields\cite{KKP04}.
Because we have performed our analysis for all $U$, we see that this remains
true at all correlation strengths, for all $\e < \mu < \e +U$, i.e., in the
region between the Coulomb peaks.  Thus there is always an opposing field
in this region, which can be understood as the origin of Coulomb blockade in terms
of density functionals.  Finally, we note that outside this region, we see regions
where the XC bias has the same sign as the applied bias.  We suspect this is an
artifact of the simplicity of the model.  Either the Anderson junction itself is
oversimplified so its results cannot be trusted in this region, or possibly this
is the non-applicability of our simple model for the spectral function [Eq. \eqref{gom}] in
this regime.  In either case, this effect should be treated with caution unless it
is also seen in a real-space calculation.

\section{Finite temperature effects}
\label{fte}

The primary focus of our work has been the low-temperature limit, because of its
relevance to standard electronic structure calculations.  However, it is relatively straightforward
to repeat our calculations at higher temperatures, as was done in Ref. \cite{KS13}. To 
do that, Eq. \eqref{den} needs to be modified as:
\ben
n=2\int_{-\infty}^{\infty} \frac{d\omega}{2\pi}f(\omega)A(\omega),
\label{nfinite}
\een
where $f(\omega)$ is the Fermi function: $f(\omega)=1/(1+\exp[(\omega-\mu)/\tau])$, with $\tau$ 
being the temperature. Also, the finite 
temperature version of Eq. \eqref{ext} needs 
to be used, i.e.,
\ben
T=-\frac{\Gamma}{2}\int_{-\infty}^{\infty}\frac{d\omega}{2\pi}\frac{df}{d\omega}A(\omega).
\label{tfinite}
\een
At finite temperature, we have no simple closed-form parametrization for the density or HXC
potential and everything we show in this section is numerical.  
For ease of comparison, we use the same temperatures as those of Ref. \cite{KS13},
namely $\tau=0.1,0.2,$ and 1.0, respectively, with $u=10$. 
We also show our low-temperature limit result.

\begin{figure}[htb]
\begin{center}
\includegraphics[width=3.5in]{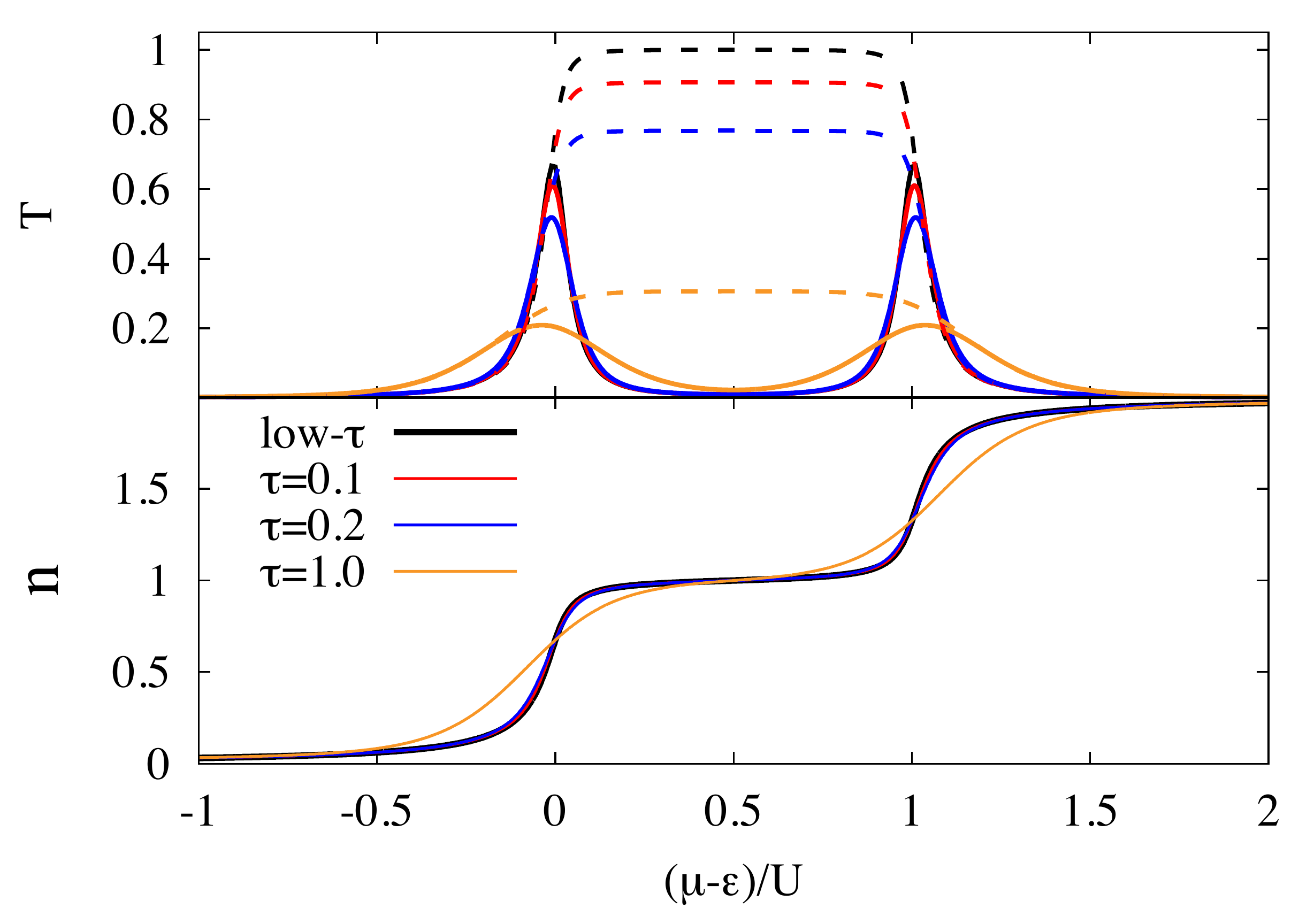}
\caption{Upper panel: conductance $T$ as a function of $(\mu-\e)/U$, plotted in unit
of $G_0$, for different temperatures $\tau$ at a fixed $u=U/\Gamma=10$. Solid lines 
are ``exact'' and dashed lines are (uncorrected) KS conductance. The black line is for
the low-temperature limit and is the same as the red line in Fig. \ref{fig1}. 
Lower panel: $n$
as a function of $(\mu-\e)/U$. By definition, the KS occupation always matches that of the 
exact occupation at all temperature.}
\label{finite}
\end{center}
\end{figure}
In Fig. \ref{finite}, we show conductance and occupation of the impurity site as a
function of $(\mu-\e)/U$ for different temperatures 
(similar results can be found in Fig. 1 of Ref. \cite{KS13}).
In the lower panel, we see that for temperatures below about 0.2, our low-temperature
limit result is indistinguishable from the numerical result.
On the other hand,  the conductance shows substantially greater 
sensitivity, both exactly and at the KS level, due to the presence of the Fermi function
in Eq. \eqref{tfinite}. 
The Friedel-Langreth sum rule applies only in the low-temperature limit. However, 
the figure also shows that our low-temperature result is approached as the temperature
is reduced.

\begin{figure}[htb]
\begin{center}
\includegraphics[width=3.5in]{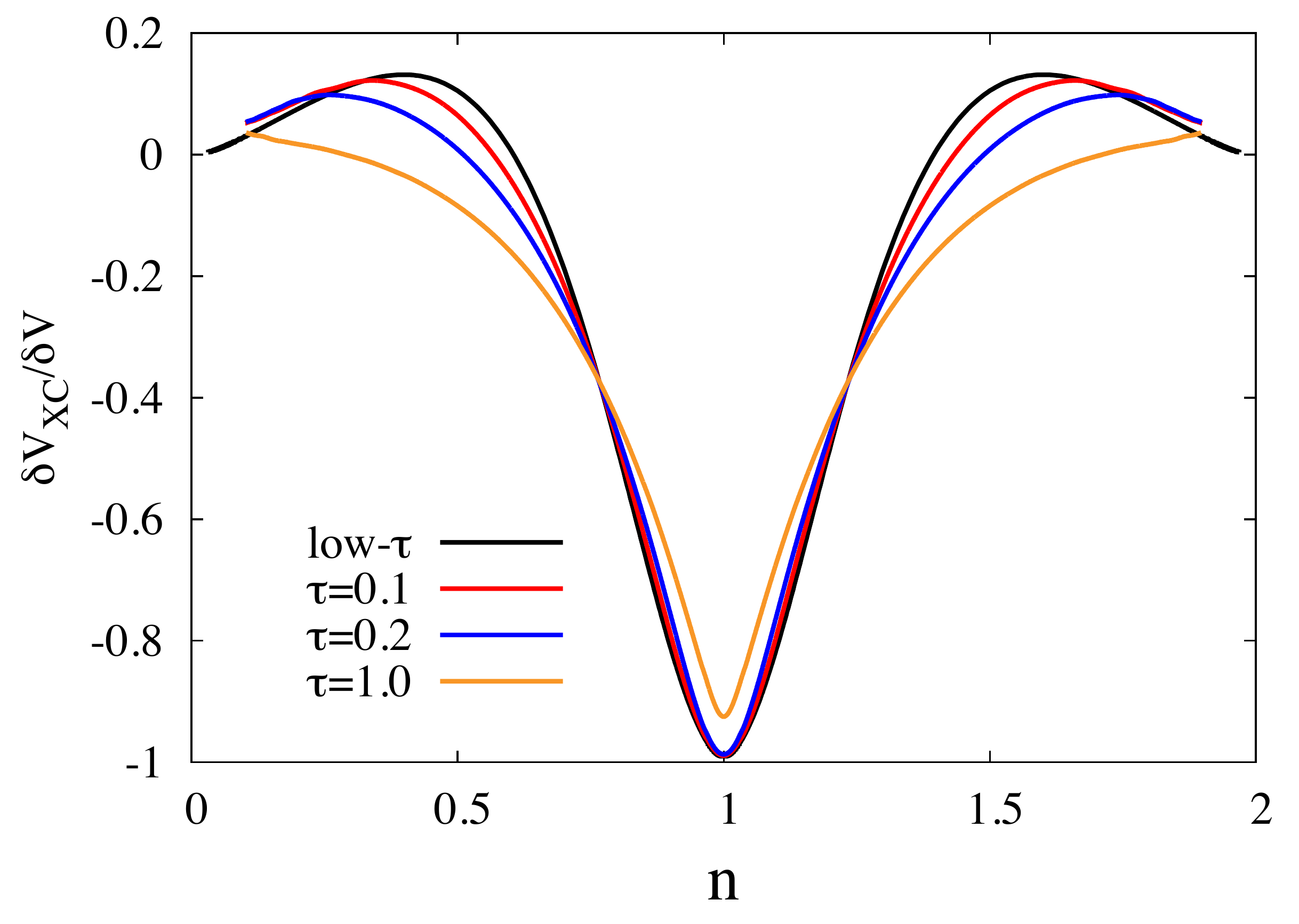}
\caption{Finite temperature generalization of Fig. \ref{dvsdv} (but with
$\delta V\xc$ plotted rather than $\delta V\s$):
$\delta V\xc$ as a function of $n$, for different temperatures $\tau$ 
at a fixed $u=U/\Gamma=10$. Black line is the low-temperature limit result.}
\label{finitedrop}
\end{center}
\end{figure}
Finally, in Fig. \ref{finitedrop} we show the XC drop $\delta V\xc/\delta V$ as a 
function of $n$, for different temperatures. One can
see that large temperatures wash out the sharp effects of the derivative
discontinuity in the strongly correlated limit, as expected.
The XC drop is less sensitive to temperature than the conductance.
Overall, the qualitative features of the XC counterfield remain, but are
weakened by increasing temperature, as the derivative discontinuity is 
rounded off.

\section{Concluding Remarks}

We conclude with a discussion of the relevance of these results for DFT
calculations of transport.  First we note that we have provided 
accurate parametrizations for both the onsite potential and the XC bias
for the junction at low (but above \tk) temperatures, that apply for
all junctions, not just those strongly coupled to the leads.
These are available for any future calculations of impurity models of
transport.  If the parameters ($\Gamma$, $U$, $\epsilon$) can be
extracted from standard ground-state DFT calculations,
the results could be compared with the accompanying DFT transport
calculations to identify limitations of the pure DFT approximations.  If $U/\Gamma$
is large, the standard DFT approach will often yield a large overestimate
of the conductance, as is often seen in comparison with experiment\cite{KCBC08}.

But, possibly  more importantly, we have suggested that an alternative origin
of the corrections to the KS conductance is non-local static XC effects.
This is very important for understanding the limitations of standard DFT
calculations of conductance and finding ways to improve them.  Our reading
suggests that, rather than looking to TDCDFT for corrections to the conductance,
one could look instead at orbital-dependent
{\em ground-state} approximations\cite{KKP04,KMK08,PSB08,RPCS08,RPC08}.
Such calculations can yield two important improvements in calculations in 
this area: (i) the correct positioning of orbital energies in the molecular
region relative to the leads, as is already well-known\cite{QVCL07}, and (ii) a finite
XC bias drop that corrects the KS conductance.

Unfortunately, due to the structureless nature of the leads in the Anderson
model, there is no way at present to distinguish between the interpretation
of the XC corrections presented in Ref. \cite{KS13} and those presented here.  
In a more realistic model, the charge is depleted from the end of one
lead and is increased at the end of the other, producing a net dipole.  Such 
a dipole contributes to the net bias drop in the Hartree potential.  But,
with an orbital-dependent functional, a counteracting effect occurs in the
exchange bias.  This can be seen mostly clearly when the molecule is only
weakly coupled to the leads.  Then the form of this counteracting field is
steps that are proportional to the applied field but that inhibit charge
from moving from the molecule to the leads.  Such steps are well-known in
the absence of an applied field, between two distinct species, where they
ensure dissociation into charge neutral fragments, and are consequences
of the derivative discontinuity \cite{PPLB82}.  Our steps are field-induced versions
of the same thing.  Only a more detailed calculation, 
such as those of Refs. \cite{SE08} and \cite{ES13}, can yield such
information.  But at least we have provided a reason to look for such corrections
when non-local ground-state approximations are being applied.

To conclude, we have shown that, in the case of the Anderson junction, 
simple parametrizations of both the onsite HXC potential and the XC bias
in linear response can reproduce the low-temperature Coulomb blockade, for all correlation
strengths, not just the strongly correlated regime discussed in Ref. \cite{KS13}.
Furthermore, the XC bias can be interpreted as a non-local
XC response to an applied electric field, and is not necessarily a dynamic
TDCDFT effect.  Qualitatively similar effects should occur in more realistic
descriptions of molecular transport\cite{QVCL07}.  
Our approximate formulas could prove useful in other contexts, or as a
check on XC approximations applied to transport problems.

\section{Acknowledgement}

We thank Justin Smith, Klaus Capelle, and Ferdinand Evers for helpful discussions. 
Work at Berkeley Lab was supported by the U.S. Department of Energy, 
Office of Basic Energy Sciences, Materials Sciences and Engineering Division, 
under Contract \# DE-AC02-05CH11231, as well as by the Molecular Foundry 
through the U.S. Department of Energy, Office of Basic Energy Sciences 
under the same contract number.
Work at UC Irvine was supported by
the U.S Department of Energy (DOE), Office of Science,
Basic Energy Sciences (BES) under award \# DE-FG02-08ER46496.

\bibliography{lit_xcdrop}

\begin{thebibliography}{56}%
\makeatletter
\providecommand \@ifxundefined [1]{%
 \@ifx{#1\undefined}
}%
\providecommand \@ifnum [1]{%
 \ifnum #1\expandafter \@firstoftwo
 \else \expandafter \@secondoftwo
 \fi
}%
\providecommand \@ifx [1]{%
 \ifx #1\expandafter \@firstoftwo
 \else \expandafter \@secondoftwo
 \fi
}%
\providecommand \natexlab [1]{#1}%
\providecommand \enquote  [1]{``#1''}%
\providecommand \bibnamefont  [1]{#1}%
\providecommand \bibfnamefont [1]{#1}%
\providecommand \citenamefont [1]{#1}%
\providecommand \href@noop [0]{\@secondoftwo}%
\providecommand \href [0]{\begingroup \@sanitize@url \@href}%
\providecommand \@href[1]{\@@startlink{#1}\@@href}%
\providecommand \@@href[1]{\endgroup#1\@@endlink}%
\providecommand \@sanitize@url [0]{\catcode `\\12\catcode `\$12\catcode
  `\&12\catcode `\#12\catcode `\^12\catcode `\_12\catcode `\%12\relax}%
\providecommand \@@startlink[1]{}%
\providecommand \@@endlink[0]{}%
\providecommand \url  [0]{\begingroup\@sanitize@url \@url }%
\providecommand \@url [1]{\endgroup\@href {#1}{\urlprefix }}%
\providecommand \urlprefix  [0]{URL }%
\providecommand \Eprint [0]{\href }%
\providecommand \doibase [0]{http://dx.doi.org/}%
\providecommand \selectlanguage [0]{\@gobble}%
\providecommand \bibinfo  [0]{\@secondoftwo}%
\providecommand \bibfield  [0]{\@secondoftwo}%
\providecommand \translation [1]{[#1]}%
\providecommand \BibitemOpen [0]{}%
\providecommand \bibitemStop [0]{}%
\providecommand \bibitemNoStop [0]{.\EOS\space}%
\providecommand \EOS [0]{\spacefactor3000\relax}%
\providecommand \BibitemShut  [1]{\csname bibitem#1\endcsname}%
\let\auto@bib@innerbib\@empty
\bibitem [{\citenamefont {Haug}\ and\ \citenamefont {Jauho}(2007)}]{HJ07}%
  \BibitemOpen
  \bibfield  {author} {\bibinfo {author} {\bibfnamefont {H.~J.~W.}\
  \bibnamefont {Haug}}\ and\ \bibinfo {author} {\bibfnamefont {A.-P.}\
  \bibnamefont {Jauho}},\ }\href@noop {} {\emph {\bibinfo {title} {Quantum
  Kinetics in Transport and Optics of Semiconductors}}},\ \bibinfo {edition}
  {2nd}\ ed.\ (\bibinfo  {publisher} {Springer},\ \bibinfo {address} {Berlin},\
  \bibinfo {year} {2007})\BibitemShut {NoStop}%
\bibitem [{\citenamefont {Meir}\ and\ \citenamefont {Wingreen}(1992)}]{MW92}%
  \BibitemOpen
  \bibfield  {author} {\bibinfo {author} {\bibfnamefont {Y.}~\bibnamefont
  {Meir}}\ and\ \bibinfo {author} {\bibfnamefont {N.~S.}\ \bibnamefont
  {Wingreen}},\ }\href {\doibase 10.1103/PhysRevLett.68.2512} {\bibfield
  {journal} {\bibinfo  {journal} {Phys. Rev. Lett.}\ }\textbf {\bibinfo
  {volume} {68}},\ \bibinfo {pages} {2512} (\bibinfo {year}
  {1992})}\BibitemShut {NoStop}%
\bibitem [{\citenamefont {Fisher}\ and\ \citenamefont {Lee}(1981)}]{FL81}%
  \BibitemOpen
  \bibfield  {author} {\bibinfo {author} {\bibfnamefont {D.~S.}\ \bibnamefont
  {Fisher}}\ and\ \bibinfo {author} {\bibfnamefont {P.~A.}\ \bibnamefont
  {Lee}},\ }\href@noop {} {\bibfield  {journal} {\bibinfo  {journal} {Phys.
  Rev. B}\ }\textbf {\bibinfo {volume} {23}},\ \bibinfo {pages} {6851}
  (\bibinfo {year} {1981})}\BibitemShut {NoStop}%
\bibitem [{\citenamefont {Taylor}\ \emph {et~al.}(2001)\citenamefont {Taylor},
  \citenamefont {Guo},\ and\ \citenamefont {Wang}}]{TGW01}%
  \BibitemOpen
  \bibfield  {author} {\bibinfo {author} {\bibfnamefont {J.}~\bibnamefont
  {Taylor}}, \bibinfo {author} {\bibfnamefont {H.}~\bibnamefont {Guo}}, \ and\
  \bibinfo {author} {\bibfnamefont {J.}~\bibnamefont {Wang}},\ }\href {\doibase
  10.1103/PhysRevB.63.245407} {\bibfield  {journal} {\bibinfo  {journal} {Phys.
  Rev. B}\ }\textbf {\bibinfo {volume} {63}},\ \bibinfo {pages} {245407}
  (\bibinfo {year} {2001})}\BibitemShut {NoStop}%
\bibitem [{\citenamefont {Brandbyge}\ \emph {et~al.}(2002)\citenamefont
  {Brandbyge}, \citenamefont {Mozos}, \citenamefont {Ordej{\'o}n},
  \citenamefont {Taylor},\ and\ \citenamefont {Stokbro}}]{transiesta}%
  \BibitemOpen
  \bibfield  {author} {\bibinfo {author} {\bibfnamefont {M.}~\bibnamefont
  {Brandbyge}}, \bibinfo {author} {\bibfnamefont {J.-L.}\ \bibnamefont
  {Mozos}}, \bibinfo {author} {\bibfnamefont {P.}~\bibnamefont {Ordej{\'o}n}},
  \bibinfo {author} {\bibfnamefont {J.}~\bibnamefont {Taylor}}, \ and\ \bibinfo
  {author} {\bibfnamefont {K.}~\bibnamefont {Stokbro}},\ }\href@noop {}
  {\bibfield  {journal} {\bibinfo  {journal} {Phys. Rev. B}\ }\textbf {\bibinfo
  {volume} {65}},\ \bibinfo {pages} {165401} (\bibinfo {year}
  {2002})}\BibitemShut {NoStop}%
\bibitem [{\citenamefont {Kohn}\ and\ \citenamefont {Sham}(1965)}]{KS65}%
  \BibitemOpen
  \bibfield  {author} {\bibinfo {author} {\bibfnamefont {W.}~\bibnamefont
  {Kohn}}\ and\ \bibinfo {author} {\bibfnamefont {L.~J.}\ \bibnamefont
  {Sham}},\ }\href {\doibase 10.1103/PhysRev.140.A1133} {\bibfield  {journal}
  {\bibinfo  {journal} {Phys. Rev.}\ }\textbf {\bibinfo {volume} {140}},\
  \bibinfo {pages} {A1133} (\bibinfo {year} {1965})}\BibitemShut {NoStop}%
\bibitem [{\citenamefont {Burke}(2012)}]{B12}%
  \BibitemOpen
  \bibfield  {author} {\bibinfo {author} {\bibfnamefont {K.}~\bibnamefont
  {Burke}},\ }\href@noop {} {\bibfield  {journal} {\bibinfo  {journal} {J.
  Chem. Phys.}\ }\textbf {\bibinfo {volume} {136}},\ \bibinfo {pages} {150901}
  (\bibinfo {year} {2012})}\BibitemShut {NoStop}%
\bibitem [{\citenamefont {Koentopp}\ \emph {et~al.}(2008)\citenamefont
  {Koentopp}, \citenamefont {Chang}, \citenamefont {Burke},\ and\ \citenamefont
  {Car}}]{KCBC08}%
  \BibitemOpen
  \bibfield  {author} {\bibinfo {author} {\bibfnamefont {M.}~\bibnamefont
  {Koentopp}}, \bibinfo {author} {\bibfnamefont {C.}~\bibnamefont {Chang}},
  \bibinfo {author} {\bibfnamefont {K.}~\bibnamefont {Burke}}, \ and\ \bibinfo
  {author} {\bibfnamefont {R.}~\bibnamefont {Car}},\ }\href@noop {} {\bibfield
  {journal} {\bibinfo  {journal} {J. Phys.: Condens. Matter}\ }\textbf
  {\bibinfo {volume} {20}},\ \bibinfo {pages} {083203} (\bibinfo {year}
  {2008})}\BibitemShut {NoStop}%
\bibitem [{\citenamefont {Toher}\ \emph {et~al.}(2005)\citenamefont {Toher},
  \citenamefont {Filippetti}, \citenamefont {Sanvito},\ and\ \citenamefont
  {Burke}}]{TFSB05}%
  \BibitemOpen
  \bibfield  {author} {\bibinfo {author} {\bibfnamefont {C.}~\bibnamefont
  {Toher}}, \bibinfo {author} {\bibfnamefont {A.}~\bibnamefont {Filippetti}},
  \bibinfo {author} {\bibfnamefont {S.}~\bibnamefont {Sanvito}}, \ and\
  \bibinfo {author} {\bibfnamefont {K.}~\bibnamefont {Burke}},\ }\href@noop {}
  {\bibfield  {journal} {\bibinfo  {journal} {Phys. Rev. Lett.}\ }\textbf
  {\bibinfo {volume} {95}},\ \bibinfo {pages} {146402} (\bibinfo {year}
  {2005})}\BibitemShut {NoStop}%
\bibitem [{\citenamefont {Anderson}(1961)}]{A61}%
  \BibitemOpen
  \bibfield  {author} {\bibinfo {author} {\bibfnamefont {P.~W.}\ \bibnamefont
  {Anderson}},\ }\href@noop {} {\bibfield  {journal} {\bibinfo  {journal}
  {Phys. Rev.}\ }\textbf {\bibinfo {volume} {124}},\ \bibinfo {pages} {41}
  (\bibinfo {year} {1961})}\BibitemShut {NoStop}%
\bibitem [{\citenamefont {Stefanucci}\ and\ \citenamefont
  {Kurth}(2011)}]{SK11}%
  \BibitemOpen
  \bibfield  {author} {\bibinfo {author} {\bibfnamefont {G.}~\bibnamefont
  {Stefanucci}}\ and\ \bibinfo {author} {\bibfnamefont {S.}~\bibnamefont
  {Kurth}},\ }\href@noop {} {\bibfield  {journal} {\bibinfo  {journal} {Phys.
  Rev. Lett.}\ }\textbf {\bibinfo {volume} {107}},\ \bibinfo {pages} {216401}
  (\bibinfo {year} {2011})}\BibitemShut {NoStop}%
\bibitem [{\citenamefont {Bergfield}\ \emph {et~al.}(2012)\citenamefont
  {Bergfield}, \citenamefont {Liu}, \citenamefont {Burke},\ and\ \citenamefont
  {Stafford}}]{BLBS12}%
  \BibitemOpen
  \bibfield  {author} {\bibinfo {author} {\bibfnamefont {J.~P.}\ \bibnamefont
  {Bergfield}}, \bibinfo {author} {\bibfnamefont {Z.-F.}\ \bibnamefont {Liu}},
  \bibinfo {author} {\bibfnamefont {K.}~\bibnamefont {Burke}}, \ and\ \bibinfo
  {author} {\bibfnamefont {C.~A.}\ \bibnamefont {Stafford}},\ }\href@noop {}
  {\bibfield  {journal} {\bibinfo  {journal} {Phys. Rev. Lett.}\ }\textbf
  {\bibinfo {volume} {108}},\ \bibinfo {pages} {066801} (\bibinfo {year}
  {2012})}\BibitemShut {NoStop}%
\bibitem [{\citenamefont {Liu}\ \emph {et~al.}(2012)\citenamefont {Liu},
  \citenamefont {Bergfield}, \citenamefont {Burke},\ and\ \citenamefont
  {Stafford}}]{LBBS12}%
  \BibitemOpen
  \bibfield  {author} {\bibinfo {author} {\bibfnamefont {Z.-F.}\ \bibnamefont
  {Liu}}, \bibinfo {author} {\bibfnamefont {J.~P.}\ \bibnamefont {Bergfield}},
  \bibinfo {author} {\bibfnamefont {K.}~\bibnamefont {Burke}}, \ and\ \bibinfo
  {author} {\bibfnamefont {C.~A.}\ \bibnamefont {Stafford}},\ }\href@noop {}
  {\bibfield  {journal} {\bibinfo  {journal} {Phys. Rev. B}\ }\textbf {\bibinfo
  {volume} {85}},\ \bibinfo {pages} {155117} (\bibinfo {year}
  {2012})}\BibitemShut {NoStop}%
\bibitem [{\citenamefont {Tr\"oster}\ \emph {et~al.}(2012)\citenamefont
  {Tr\"oster}, \citenamefont {Schmitteckert},\ and\ \citenamefont
  {Evers}}]{TSE12}%
  \BibitemOpen
  \bibfield  {author} {\bibinfo {author} {\bibfnamefont {P.}~\bibnamefont
  {Tr\"oster}}, \bibinfo {author} {\bibfnamefont {P.}~\bibnamefont
  {Schmitteckert}}, \ and\ \bibinfo {author} {\bibfnamefont {F.}~\bibnamefont
  {Evers}},\ }\href {http://link.aps.org/doi/10.1103/PhysRevB.85.115409}
  {\bibfield  {journal} {\bibinfo  {journal} {Phys. Rev. B}\ }\textbf {\bibinfo
  {volume} {85}},\ \bibinfo {pages} {115409} (\bibinfo {year}
  {2012})}\BibitemShut {NoStop}%
\bibitem [{\citenamefont {Kurth}\ and\ \citenamefont
  {Stefanucci}(2013)}]{KS13}%
  \BibitemOpen
  \bibfield  {author} {\bibinfo {author} {\bibfnamefont {S.}~\bibnamefont
  {Kurth}}\ and\ \bibinfo {author} {\bibfnamefont {G.}~\bibnamefont
  {Stefanucci}},\ }\href@noop {} {\bibfield  {journal} {\bibinfo  {journal}
  {Phys. Rev. Lett.}\ }\textbf {\bibinfo {volume} {111}},\ \bibinfo {pages}
  {030601} (\bibinfo {year} {2013})}\BibitemShut {NoStop}%
\bibitem [{\citenamefont {Stefanucci}\ and\ \citenamefont
  {Kurth}(2013)}]{SK13}%
  \BibitemOpen
  \bibfield  {author} {\bibinfo {author} {\bibfnamefont {G.}~\bibnamefont
  {Stefanucci}}\ and\ \bibinfo {author} {\bibfnamefont {S.}~\bibnamefont
  {Kurth}},\ }\href@noop {} {\bibfield  {journal} {\bibinfo  {journal} {Phys.
  Status Solidi B}\ }\textbf {\bibinfo {volume} {250}},\ \bibinfo {pages}
  {2378} (\bibinfo {year} {2013})}\BibitemShut {NoStop}%
\bibitem [{\citenamefont {Evers}\ and\ \citenamefont
  {Schmitteckert}(2013{\natexlab{a}})}]{ES13}%
  \BibitemOpen
  \bibfield  {author} {\bibinfo {author} {\bibfnamefont {F.}~\bibnamefont
  {Evers}}\ and\ \bibinfo {author} {\bibfnamefont {P.}~\bibnamefont
  {Schmitteckert}},\ }\href@noop {} {\bibfield  {journal} {\bibinfo  {journal}
  {Phys. Status Solidi B}\ }\textbf {\bibinfo {volume} {250}},\ \bibinfo
  {pages} {2330} (\bibinfo {year} {2013}{\natexlab{a}})}\BibitemShut {NoStop}%
\bibitem [{\citenamefont {Evers}\ and\ \citenamefont
  {Schmitteckert}(2013{\natexlab{b}})}]{ES13b}%
  \BibitemOpen
  \bibfield  {author} {\bibinfo {author} {\bibfnamefont {F.}~\bibnamefont
  {Evers}}\ and\ \bibinfo {author} {\bibfnamefont {P.}~\bibnamefont
  {Schmitteckert}},\ }\href@noop {} {\bibfield  {journal} {\bibinfo  {journal}
  {Eur. Phys. Lett.}\ }\textbf {\bibinfo {volume} {103}},\ \bibinfo {pages}
  {47012} (\bibinfo {year} {2013}{\natexlab{b}})}\BibitemShut {NoStop}%
\bibitem [{\citenamefont {Wiegmann}\ and\ \citenamefont
  {Tsvelick}(1983)}]{WT83}%
  \BibitemOpen
  \bibfield  {author} {\bibinfo {author} {\bibfnamefont {P.~B.}\ \bibnamefont
  {Wiegmann}}\ and\ \bibinfo {author} {\bibfnamefont {A.~M.}\ \bibnamefont
  {Tsvelick}},\ }\href {http://iopscience.iop.org/0022-3719/16/12/017}
  {\bibfield  {journal} {\bibinfo  {journal} {J. Phys. C: Solid State Phys.}\
  }\textbf {\bibinfo {volume} {16}},\ \bibinfo {pages} {2281} (\bibinfo {year}
  {1983})}\BibitemShut {NoStop}%
\bibitem [{\citenamefont {Friedel}(1958)}]{F58}%
  \BibitemOpen
  \bibfield  {author} {\bibinfo {author} {\bibfnamefont {J.}~\bibnamefont
  {Friedel}},\ }\href@noop {} {\bibfield  {journal} {\bibinfo  {journal} {Nuovo
  Cimento Suppl}\ }\textbf {\bibinfo {volume} {7}},\ \bibinfo {pages} {287}
  (\bibinfo {year} {1958})}\BibitemShut {NoStop}%
\bibitem [{\citenamefont {Langreth}(1966)}]{L66}%
  \BibitemOpen
  \bibfield  {author} {\bibinfo {author} {\bibfnamefont {D.~C.}\ \bibnamefont
  {Langreth}},\ }\href@noop {} {\bibfield  {journal} {\bibinfo  {journal}
  {Phys. Rev.}\ }\textbf {\bibinfo {volume} {150}},\ \bibinfo {pages} {516}
  (\bibinfo {year} {1966})}\BibitemShut {NoStop}%
\bibitem [{\citenamefont {Perdew}\ \emph {et~al.}(1982)\citenamefont {Perdew},
  \citenamefont {Parr}, \citenamefont {Levy},\ and\ \citenamefont
  {Balduz}}]{PPLB82}%
  \BibitemOpen
  \bibfield  {author} {\bibinfo {author} {\bibfnamefont {J.~P.}\ \bibnamefont
  {Perdew}}, \bibinfo {author} {\bibfnamefont {R.~G.}\ \bibnamefont {Parr}},
  \bibinfo {author} {\bibfnamefont {M.}~\bibnamefont {Levy}}, \ and\ \bibinfo
  {author} {\bibfnamefont {J.~L.}\ \bibnamefont {Balduz}},\ }\href@noop {}
  {\bibfield  {journal} {\bibinfo  {journal} {Phys. Rev. Lett.}\ }\textbf
  {\bibinfo {volume} {49}},\ \bibinfo {pages} {1691} (\bibinfo {year}
  {1982})}\BibitemShut {NoStop}%
\bibitem [{\citenamefont {Capelle}\ \emph {et~al.}(2007)\citenamefont
  {Capelle}, \citenamefont {Borgh}, \citenamefont {K\"{a}rkk\"{a}inen},\ and\
  \citenamefont {Reimann}}]{CBKR07}%
  \BibitemOpen
  \bibfield  {author} {\bibinfo {author} {\bibfnamefont {K.}~\bibnamefont
  {Capelle}}, \bibinfo {author} {\bibfnamefont {M.}~\bibnamefont {Borgh}},
  \bibinfo {author} {\bibfnamefont {K.}~\bibnamefont {K\"{a}rkk\"{a}inen}}, \
  and\ \bibinfo {author} {\bibfnamefont {S.~M.}\ \bibnamefont {Reimann}},\
  }\href@noop {} {\bibfield  {journal} {\bibinfo  {journal} {Phys. Rev. Lett.}\
  }\textbf {\bibinfo {volume} {99}},\ \bibinfo {pages} {010402} (\bibinfo
  {year} {2007})}\BibitemShut {NoStop}%
\bibitem [{\citenamefont {Kurth}\ \emph {et~al.}(2010)\citenamefont {Kurth},
  \citenamefont {Stefanucci}, \citenamefont {Khosravi}, \citenamefont
  {Verdozzi},\ and\ \citenamefont {Gross}}]{KSKV10}%
  \BibitemOpen
  \bibfield  {author} {\bibinfo {author} {\bibfnamefont {S.}~\bibnamefont
  {Kurth}}, \bibinfo {author} {\bibfnamefont {G.}~\bibnamefont {Stefanucci}},
  \bibinfo {author} {\bibfnamefont {E.}~\bibnamefont {Khosravi}}, \bibinfo
  {author} {\bibfnamefont {C.}~\bibnamefont {Verdozzi}}, \ and\ \bibinfo
  {author} {\bibfnamefont {E.~K.~U.}\ \bibnamefont {Gross}},\ }\href@noop {}
  {\bibfield  {journal} {\bibinfo  {journal} {Phys. Rev. Lett.}\ }\textbf
  {\bibinfo {volume} {104}},\ \bibinfo {pages} {236801} (\bibinfo {year}
  {2010})}\BibitemShut {NoStop}%
\bibitem [{\citenamefont {Pustilnik}\ and\ \citenamefont
  {Glazman}(2004)}]{PG04}%
  \BibitemOpen
  \bibfield  {author} {\bibinfo {author} {\bibfnamefont {M.}~\bibnamefont
  {Pustilnik}}\ and\ \bibinfo {author} {\bibfnamefont {L.}~\bibnamefont
  {Glazman}},\ }\href@noop {} {\bibfield  {journal} {\bibinfo  {journal} {J.
  Phys.: Condens. Matter}\ }\textbf {\bibinfo {volume} {16}},\ \bibinfo {pages}
  {R513} (\bibinfo {year} {2004})}\BibitemShut {NoStop}%
\bibitem [{not({\natexlab{a}})}]{note2}%
  \BibitemOpen
  \href@noop {} {}\bibinfo {howpublished} {S. Kurth and G. Stefanucci, private
  communication.} ({\natexlab{a}})\BibitemShut {NoStop}%
\bibitem [{\citenamefont {Sai}\ \emph {et~al.}(2005)\citenamefont {Sai},
  \citenamefont {Zwolak}, \citenamefont {Vignale},\ and\ \citenamefont
  {Di~Ventra}}]{SZVD05}%
  \BibitemOpen
  \bibfield  {author} {\bibinfo {author} {\bibfnamefont {N.}~\bibnamefont
  {Sai}}, \bibinfo {author} {\bibfnamefont {M.}~\bibnamefont {Zwolak}},
  \bibinfo {author} {\bibfnamefont {G.}~\bibnamefont {Vignale}}, \ and\
  \bibinfo {author} {\bibfnamefont {M.}~\bibnamefont {Di~Ventra}},\ }\href@noop
  {} {\bibfield  {journal} {\bibinfo  {journal} {Phys. Rev. Lett.}\ }\textbf
  {\bibinfo {volume} {94}},\ \bibinfo {pages} {186810} (\bibinfo {year}
  {2005})}\BibitemShut {NoStop}%
\bibitem [{\citenamefont {Koentopp}\ \emph {et~al.}(2006)\citenamefont
  {Koentopp}, \citenamefont {Burke},\ and\ \citenamefont {Evers}}]{KBE06}%
  \BibitemOpen
  \bibfield  {author} {\bibinfo {author} {\bibfnamefont {M.}~\bibnamefont
  {Koentopp}}, \bibinfo {author} {\bibfnamefont {K.}~\bibnamefont {Burke}}, \
  and\ \bibinfo {author} {\bibfnamefont {F.}~\bibnamefont {Evers}},\
  }\href@noop {} {\bibfield  {journal} {\bibinfo  {journal} {Phys. Rev. B}\
  }\textbf {\bibinfo {volume} {73}},\ \bibinfo {pages} {121403} (\bibinfo
  {year} {2006})}\BibitemShut {NoStop}%
\bibitem [{\citenamefont {Vignale}\ and\ \citenamefont
  {Di~Ventra}(2009)}]{VD09}%
  \BibitemOpen
  \bibfield  {author} {\bibinfo {author} {\bibfnamefont {G.}~\bibnamefont
  {Vignale}}\ and\ \bibinfo {author} {\bibfnamefont {M.}~\bibnamefont
  {Di~Ventra}},\ }\href@noop {} {\bibfield  {journal} {\bibinfo  {journal}
  {Phys. Rev. B}\ }\textbf {\bibinfo {volume} {79}},\ \bibinfo {pages} {014201}
  (\bibinfo {year} {2009})}\BibitemShut {NoStop}%
\bibitem [{\citenamefont {K\"{u}mmel}\ \emph {et~al.}(2004)\citenamefont
  {K\"{u}mmel}, \citenamefont {Kronik},\ and\ \citenamefont {Perdew}}]{KKP04}%
  \BibitemOpen
  \bibfield  {author} {\bibinfo {author} {\bibfnamefont {S.}~\bibnamefont
  {K\"{u}mmel}}, \bibinfo {author} {\bibfnamefont {L.}~\bibnamefont {Kronik}},
  \ and\ \bibinfo {author} {\bibfnamefont {J.~P.}\ \bibnamefont {Perdew}},\
  }\href@noop {} {\bibfield  {journal} {\bibinfo  {journal} {Phys. Rev. Lett.}\
  }\textbf {\bibinfo {volume} {93}},\ \bibinfo {pages} {213002} (\bibinfo
  {year} {2004})}\BibitemShut {NoStop}%
\bibitem [{\citenamefont {K\"{u}mmel}\ and\ \citenamefont
  {Kronik}(2008)}]{KK08}%
  \BibitemOpen
  \bibfield  {author} {\bibinfo {author} {\bibfnamefont {S.}~\bibnamefont
  {K\"{u}mmel}}\ and\ \bibinfo {author} {\bibfnamefont {L.}~\bibnamefont
  {Kronik}},\ }\href@noop {} {\bibfield  {journal} {\bibinfo  {journal} {Rev.
  Mod. Phys.}\ }\textbf {\bibinfo {volume} {80}},\ \bibinfo {pages} {3}
  (\bibinfo {year} {2008})}\BibitemShut {NoStop}%
\bibitem [{not({\natexlab{b}})}]{note1}%
  \BibitemOpen
  \href@noop {} {}\bibinfo {howpublished} {In Refs. \cite{BLBS12} and
  \cite{LBBS12}, we used $\Gamma=\Gamma_L=\Gamma_R$. Here, we use
  $\Gamma=\Gamma_L+\Gamma_R$, consistent with Ref. \cite{KS13}, which is more
  relevant to the discussions in this work.} ({\natexlab{b}})\BibitemShut
  {NoStop}%
\bibitem [{\citenamefont {Silver}\ \emph {et~al.}(1990)\citenamefont {Silver},
  \citenamefont {Gubernatis}, \citenamefont {Sivia},\ and\ \citenamefont
  {Jarrell}}]{SGSJ90}%
  \BibitemOpen
  \bibfield  {author} {\bibinfo {author} {\bibfnamefont {R.~N.}\ \bibnamefont
  {Silver}}, \bibinfo {author} {\bibfnamefont {J.~E.}\ \bibnamefont
  {Gubernatis}}, \bibinfo {author} {\bibfnamefont {D.~S.}\ \bibnamefont
  {Sivia}}, \ and\ \bibinfo {author} {\bibfnamefont {M.}~\bibnamefont
  {Jarrell}},\ }\href {http://prl.aps.org/abstract/PRL/v65/i4/p496_1}
  {\bibfield  {journal} {\bibinfo  {journal} {Phys. Rev. Lett.}\ }\textbf
  {\bibinfo {volume} {65}},\ \bibinfo {pages} {496} (\bibinfo {year}
  {1990})}\BibitemShut {NoStop}%
\bibitem [{\citenamefont {Carrascal}\ \emph {et~al.}(2015)\citenamefont
  {Carrascal}, \citenamefont {Ferrer}, \citenamefont {Smith},\ and\
  \citenamefont {Burke}}]{CFSB15}%
  \BibitemOpen
  \bibfield  {author} {\bibinfo {author} {\bibfnamefont {D.}~\bibnamefont
  {Carrascal}}, \bibinfo {author} {\bibfnamefont {J.}~\bibnamefont {Ferrer}},
  \bibinfo {author} {\bibfnamefont {J.~C.}\ \bibnamefont {Smith}}, \ and\
  \bibinfo {author} {\bibfnamefont {K.}~\bibnamefont {Burke}},\ }\href@noop {}
  {\enquote {\bibinfo {title} {The {H}ubbard dimer: A density functional case
  study of a many-body problem},}\ }\bibinfo {howpublished} {ArXiv: 1502.02194}
  (\bibinfo {year} {2015})\BibitemShut {NoStop}%
\bibitem [{\citenamefont {Hewson}(1997)}]{H97}%
  \BibitemOpen
  \bibfield  {author} {\bibinfo {author} {\bibfnamefont {A.~C.}\ \bibnamefont
  {Hewson}},\ }\href@noop {} {\emph {\bibinfo {title} {The Kondo problem to
  heavy fermions}}}\ (\bibinfo  {publisher} {Cambridge University Press},\
  \bibinfo {address} {Cambridge},\ \bibinfo {year} {1997})\BibitemShut
  {NoStop}%
\bibitem [{\citenamefont {Mermin}(1965)}]{M65}%
  \BibitemOpen
  \bibfield  {author} {\bibinfo {author} {\bibfnamefont {N.~D.}\ \bibnamefont
  {Mermin}},\ }\href@noop {} {\bibfield  {journal} {\bibinfo  {journal} {Phys.
  Rev.}\ }\textbf {\bibinfo {volume} {137}},\ \bibinfo {pages} {A1441}
  (\bibinfo {year} {1965})}\BibitemShut {NoStop}%
\bibitem [{\citenamefont {Kamenev}\ and\ \citenamefont {Kohn}(2001)}]{KK01}%
  \BibitemOpen
  \bibfield  {author} {\bibinfo {author} {\bibfnamefont {A.}~\bibnamefont
  {Kamenev}}\ and\ \bibinfo {author} {\bibfnamefont {W.}~\bibnamefont {Kohn}},\
  }\href@noop {} {\bibfield  {journal} {\bibinfo  {journal} {Phys. Rev. B}\
  }\textbf {\bibinfo {volume} {63}},\ \bibinfo {pages} {155304} (\bibinfo
  {year} {2001})}\BibitemShut {NoStop}%
\bibitem [{\citenamefont {Baranger}\ and\ \citenamefont {Stone}(1989)}]{BS89}%
  \BibitemOpen
  \bibfield  {author} {\bibinfo {author} {\bibfnamefont {H.~U.}\ \bibnamefont
  {Baranger}}\ and\ \bibinfo {author} {\bibfnamefont {A.~D.}\ \bibnamefont
  {Stone}},\ }\href@noop {} {\bibfield  {journal} {\bibinfo  {journal} {Phys.
  Rev. B}\ }\textbf {\bibinfo {volume} {40}},\ \bibinfo {pages} {8169}
  (\bibinfo {year} {1989})}\BibitemShut {NoStop}%
\bibitem [{\citenamefont {Stefanucci}\ \emph {et~al.}(2006)\citenamefont
  {Stefanucci}, \citenamefont {Almbladh}, \citenamefont {Kurth}, \citenamefont
  {Gross}, \citenamefont {Rubio}, \citenamefont {van Leeuwen}, \citenamefont
  {Dahlen},\ and\ \citenamefont {von Barth}}]{SAKG06}%
  \BibitemOpen
  \bibfield  {author} {\bibinfo {author} {\bibfnamefont {G.}~\bibnamefont
  {Stefanucci}}, \bibinfo {author} {\bibfnamefont {C.-O.}\ \bibnamefont
  {Almbladh}}, \bibinfo {author} {\bibfnamefont {S.}~\bibnamefont {Kurth}},
  \bibinfo {author} {\bibfnamefont {E.~K.~U.}\ \bibnamefont {Gross}}, \bibinfo
  {author} {\bibfnamefont {A.}~\bibnamefont {Rubio}}, \bibinfo {author}
  {\bibfnamefont {R.}~\bibnamefont {van Leeuwen}}, \bibinfo {author}
  {\bibfnamefont {N.~E.}\ \bibnamefont {Dahlen}}, \ and\ \bibinfo {author}
  {\bibfnamefont {U.}~\bibnamefont {von Barth}},\ }in\ \href@noop {} {\emph
  {\bibinfo {booktitle} {Time-Dependent Density Functional Theory}}},\ \bibinfo
  {editor} {edited by\ \bibinfo {editor} {\bibfnamefont {M.~A.~L.}\
  \bibnamefont {Marques}}, \bibinfo {editor} {\bibfnamefont {C.~A.}\
  \bibnamefont {Ullrich}}, \bibinfo {editor} {\bibfnamefont {F.}~\bibnamefont
  {Nogueira}}, \bibinfo {editor} {\bibfnamefont {A.}~\bibnamefont {Rubio}},
  \bibinfo {editor} {\bibfnamefont {K.}~\bibnamefont {Burke}}, \ and\ \bibinfo
  {editor} {\bibfnamefont {E.~K.~U.}\ \bibnamefont {Gross}}}\ (\bibinfo
  {publisher} {Springer},\ \bibinfo {year} {2006})\BibitemShut {NoStop}%
\bibitem [{\citenamefont {Gonze}\ \emph {et~al.}(1995)\citenamefont {Gonze},
  \citenamefont {Ghosez},\ and\ \citenamefont {Godby}}]{GGG95}%
  \BibitemOpen
  \bibfield  {author} {\bibinfo {author} {\bibfnamefont {X.}~\bibnamefont
  {Gonze}}, \bibinfo {author} {\bibfnamefont {{\relax Ph}.}~\bibnamefont
  {Ghosez}}, \ and\ \bibinfo {author} {\bibfnamefont {R.~W.}\ \bibnamefont
  {Godby}},\ }\href@noop {} {\bibfield  {journal} {\bibinfo  {journal} {Phys.
  Rev. Lett.}\ }\textbf {\bibinfo {volume} {74}},\ \bibinfo {pages} {4035}
  (\bibinfo {year} {1995})}\BibitemShut {NoStop}%
\bibitem [{\citenamefont {Gonze}\ \emph {et~al.}(1997)\citenamefont {Gonze},
  \citenamefont {Ghosez},\ and\ \citenamefont {Godby}}]{GGG97}%
  \BibitemOpen
  \bibfield  {author} {\bibinfo {author} {\bibfnamefont {X.}~\bibnamefont
  {Gonze}}, \bibinfo {author} {\bibfnamefont {{\relax Ph}.}~\bibnamefont
  {Ghosez}}, \ and\ \bibinfo {author} {\bibfnamefont {R.~W.}\ \bibnamefont
  {Godby}},\ }\href@noop {} {\bibfield  {journal} {\bibinfo  {journal} {Phys.
  Rev. Lett.}\ }\textbf {\bibinfo {volume} {78}},\ \bibinfo {pages} {294}
  (\bibinfo {year} {1997})}\BibitemShut {NoStop}%
\bibitem [{\citenamefont {Resta}(1994)}]{R94}%
  \BibitemOpen
  \bibfield  {author} {\bibinfo {author} {\bibfnamefont {R.}~\bibnamefont
  {Resta}},\ }\href@noop {} {\bibfield  {journal} {\bibinfo  {journal} {Rev.
  Mod. Phys.}\ }\textbf {\bibinfo {volume} {66}},\ \bibinfo {pages} {899}
  (\bibinfo {year} {1994})}\BibitemShut {NoStop}%
\bibitem [{\citenamefont {Onida}\ \emph {et~al.}(2002)\citenamefont {Onida},
  \citenamefont {Reining},\ and\ \citenamefont {Rubio}}]{ORR02}%
  \BibitemOpen
  \bibfield  {author} {\bibinfo {author} {\bibfnamefont {G.}~\bibnamefont
  {Onida}}, \bibinfo {author} {\bibfnamefont {L.}~\bibnamefont {Reining}}, \
  and\ \bibinfo {author} {\bibfnamefont {A.}~\bibnamefont {Rubio}},\ }\href
  {\doibase 10.1103/RevModPhys.74.601} {\bibfield  {journal} {\bibinfo
  {journal} {Rev. Mod. Phys.}\ }\textbf {\bibinfo {volume} {74}},\ \bibinfo
  {pages} {601} (\bibinfo {year} {2002})}\BibitemShut {NoStop}%
\bibitem [{\citenamefont {Champagne}\ \emph {et~al.}(1998)\citenamefont
  {Champagne}, \citenamefont {Perp\`{e}te}, \citenamefont {van Gisbergen},
  \citenamefont {Baerends}, \citenamefont {Snijders}, \citenamefont
  {Soubra-Ghaoui}, \citenamefont {Robins},\ and\ \citenamefont
  {Kirtman}}]{CPGB98}%
  \BibitemOpen
  \bibfield  {author} {\bibinfo {author} {\bibfnamefont {B.}~\bibnamefont
  {Champagne}}, \bibinfo {author} {\bibfnamefont {E.~A.}\ \bibnamefont
  {Perp\`{e}te}}, \bibinfo {author} {\bibfnamefont {S.~J.~A.}\ \bibnamefont
  {van Gisbergen}}, \bibinfo {author} {\bibfnamefont {E.-J.}\ \bibnamefont
  {Baerends}}, \bibinfo {author} {\bibfnamefont {J.~G.}\ \bibnamefont
  {Snijders}}, \bibinfo {author} {\bibfnamefont {C.}~\bibnamefont
  {Soubra-Ghaoui}}, \bibinfo {author} {\bibfnamefont {K.~A.}\ \bibnamefont
  {Robins}}, \ and\ \bibinfo {author} {\bibfnamefont {B.}~\bibnamefont
  {Kirtman}},\ }\href@noop {} {\bibfield  {journal} {\bibinfo  {journal} {J.
  Chem. Phys.}\ }\textbf {\bibinfo {volume} {109}},\ \bibinfo {pages} {10489}
  (\bibinfo {year} {1998})}\BibitemShut {NoStop}%
\bibitem [{\citenamefont {van Gisbergen}\ \emph {et~al.}(1999)\citenamefont
  {van Gisbergen}, \citenamefont {Schipper}, \citenamefont {Gritsenko},
  \citenamefont {Baerends}, \citenamefont {Snijders}, \citenamefont
  {Champagne},\ and\ \citenamefont {Kirtman}}]{GSGB99}%
  \BibitemOpen
  \bibfield  {author} {\bibinfo {author} {\bibfnamefont {S.~J.~A.}\
  \bibnamefont {van Gisbergen}}, \bibinfo {author} {\bibfnamefont {P.~R.~T.}\
  \bibnamefont {Schipper}}, \bibinfo {author} {\bibfnamefont {O.~V.}\
  \bibnamefont {Gritsenko}}, \bibinfo {author} {\bibfnamefont {E.~J.}\
  \bibnamefont {Baerends}}, \bibinfo {author} {\bibfnamefont {J.~G.}\
  \bibnamefont {Snijders}}, \bibinfo {author} {\bibfnamefont {B.}~\bibnamefont
  {Champagne}}, \ and\ \bibinfo {author} {\bibfnamefont {B.}~\bibnamefont
  {Kirtman}},\ }\href@noop {} {\bibfield  {journal} {\bibinfo  {journal} {Phys.
  Rev. Lett.}\ }\textbf {\bibinfo {volume} {83}},\ \bibinfo {pages} {694}
  (\bibinfo {year} {1999})}\BibitemShut {NoStop}%
\bibitem [{\citenamefont {van Faassen}\ \emph {et~al.}(2002)\citenamefont {van
  Faassen}, \citenamefont {de~Boeij}, \citenamefont {van Leeuwen},
  \citenamefont {Berger},\ and\ \citenamefont {Snijders}}]{FBLB02}%
  \BibitemOpen
  \bibfield  {author} {\bibinfo {author} {\bibfnamefont {M.}~\bibnamefont {van
  Faassen}}, \bibinfo {author} {\bibfnamefont {P.~L.}\ \bibnamefont
  {de~Boeij}}, \bibinfo {author} {\bibfnamefont {R.}~\bibnamefont {van
  Leeuwen}}, \bibinfo {author} {\bibfnamefont {J.~A.}\ \bibnamefont {Berger}},
  \ and\ \bibinfo {author} {\bibfnamefont {J.~G.}\ \bibnamefont {Snijders}},\
  }\href@noop {} {\bibfield  {journal} {\bibinfo  {journal} {Phys. Rev. Lett.}\
  }\textbf {\bibinfo {volume} {88}},\ \bibinfo {pages} {186401} (\bibinfo
  {year} {2002})}\BibitemShut {NoStop}%
\bibitem [{\citenamefont {Pemmaraju}\ \emph {et~al.}(2008)\citenamefont
  {Pemmaraju}, \citenamefont {Sanvito},\ and\ \citenamefont {Burke}}]{PSB08}%
  \BibitemOpen
  \bibfield  {author} {\bibinfo {author} {\bibfnamefont {C.~D.}\ \bibnamefont
  {Pemmaraju}}, \bibinfo {author} {\bibfnamefont {S.}~\bibnamefont {Sanvito}},
  \ and\ \bibinfo {author} {\bibfnamefont {K.}~\bibnamefont {Burke}},\
  }\href@noop {} {\bibfield  {journal} {\bibinfo  {journal} {Phys. Rev. B}\
  }\textbf {\bibinfo {volume} {77}},\ \bibinfo {pages} {121204} (\bibinfo
  {year} {2008})}\BibitemShut {NoStop}%
\bibitem [{\citenamefont {Berger}\ \emph {et~al.}(2007)\citenamefont {Berger},
  \citenamefont {de~Boeij},\ and\ \citenamefont {van Leeuwen}}]{BBL07}%
  \BibitemOpen
  \bibfield  {author} {\bibinfo {author} {\bibfnamefont {J.~A.}\ \bibnamefont
  {Berger}}, \bibinfo {author} {\bibfnamefont {P.~L.}\ \bibnamefont
  {de~Boeij}}, \ and\ \bibinfo {author} {\bibfnamefont {R.}~\bibnamefont {van
  Leeuwen}},\ }\href@noop {} {\bibfield  {journal} {\bibinfo  {journal} {Phys.
  Rev. B}\ }\textbf {\bibinfo {volume} {75}},\ \bibinfo {pages} {035116}
  (\bibinfo {year} {2007})}\BibitemShut {NoStop}%
\bibitem [{\citenamefont {Jung}\ \emph {et~al.}(2007)\citenamefont {Jung},
  \citenamefont {Bokes},\ and\ \citenamefont {Godby}}]{JBG07}%
  \BibitemOpen
  \bibfield  {author} {\bibinfo {author} {\bibfnamefont {J.}~\bibnamefont
  {Jung}}, \bibinfo {author} {\bibfnamefont {P.}~\bibnamefont {Bokes}}, \ and\
  \bibinfo {author} {\bibfnamefont {R.~W.}\ \bibnamefont {Godby}},\ }\href@noop
  {} {\bibfield  {journal} {\bibinfo  {journal} {Phys. Rev. Lett.}\ }\textbf
  {\bibinfo {volume} {98}},\ \bibinfo {pages} {259701} (\bibinfo {year}
  {2007})}\BibitemShut {NoStop}%
\bibitem [{\citenamefont {Maitra}\ \emph {et~al.}(2003)\citenamefont {Maitra},
  \citenamefont {Souza},\ and\ \citenamefont {Burke}}]{MSB03}%
  \BibitemOpen
  \bibfield  {author} {\bibinfo {author} {\bibfnamefont {N.~T.}\ \bibnamefont
  {Maitra}}, \bibinfo {author} {\bibfnamefont {I.}~\bibnamefont {Souza}}, \
  and\ \bibinfo {author} {\bibfnamefont {K.}~\bibnamefont {Burke}},\
  }\href@noop {} {\bibfield  {journal} {\bibinfo  {journal} {Phys. Rev. B}\
  }\textbf {\bibinfo {volume} {68}},\ \bibinfo {pages} {045109} (\bibinfo
  {year} {2003})}\BibitemShut {NoStop}%
\bibitem [{\citenamefont {Kohn}(1964)}]{K64}%
  \BibitemOpen
  \bibfield  {author} {\bibinfo {author} {\bibfnamefont {W.}~\bibnamefont
  {Kohn}},\ }\href@noop {} {\bibfield  {journal} {\bibinfo  {journal} {Phys.
  Rev.}\ }\textbf {\bibinfo {volume} {133}},\ \bibinfo {pages} {A171} (\bibinfo
  {year} {1964})}\BibitemShut {NoStop}%
\bibitem [{\citenamefont {K\"{o}rzd\"{o}rfer}\ \emph
  {et~al.}(2008)\citenamefont {K\"{o}rzd\"{o}rfer}, \citenamefont {Mundt},\
  and\ \citenamefont {K\"{u}mmel}}]{KMK08}%
  \BibitemOpen
  \bibfield  {author} {\bibinfo {author} {\bibfnamefont {T.}~\bibnamefont
  {K\"{o}rzd\"{o}rfer}}, \bibinfo {author} {\bibfnamefont {M.}~\bibnamefont
  {Mundt}}, \ and\ \bibinfo {author} {\bibfnamefont {S.}~\bibnamefont
  {K\"{u}mmel}},\ }\href@noop {} {\bibfield  {journal} {\bibinfo  {journal}
  {Phys. Rev. Lett.}\ }\textbf {\bibinfo {volume} {100}},\ \bibinfo {pages}
  {133004} (\bibinfo {year} {2008})}\BibitemShut {NoStop}%
\bibitem [{\citenamefont {Ruzsinszky}\ \emph
  {et~al.}(2008{\natexlab{a}})\citenamefont {Ruzsinszky}, \citenamefont
  {Perdew}, \citenamefont {Csonka}, \citenamefont {Scuseria},\ and\
  \citenamefont {Vydrov}}]{RPCS08}%
  \BibitemOpen
  \bibfield  {author} {\bibinfo {author} {\bibfnamefont {A.}~\bibnamefont
  {Ruzsinszky}}, \bibinfo {author} {\bibfnamefont {J.~P.}\ \bibnamefont
  {Perdew}}, \bibinfo {author} {\bibfnamefont {G.~I.}\ \bibnamefont {Csonka}},
  \bibinfo {author} {\bibfnamefont {G.~E.}\ \bibnamefont {Scuseria}}, \ and\
  \bibinfo {author} {\bibfnamefont {O.~A.}\ \bibnamefont {Vydrov}},\
  }\href@noop {} {\bibfield  {journal} {\bibinfo  {journal} {Phys. Rev. A}\
  }\textbf {\bibinfo {volume} {77}},\ \bibinfo {pages} {060502(R)} (\bibinfo
  {year} {2008}{\natexlab{a}})}\BibitemShut {NoStop}%
\bibitem [{\citenamefont {Ruzsinszky}\ \emph
  {et~al.}(2008{\natexlab{b}})\citenamefont {Ruzsinszky}, \citenamefont
  {Perdew},\ and\ \citenamefont {Csonka}}]{RPC08}%
  \BibitemOpen
  \bibfield  {author} {\bibinfo {author} {\bibfnamefont {A.}~\bibnamefont
  {Ruzsinszky}}, \bibinfo {author} {\bibfnamefont {J.~P.}\ \bibnamefont
  {Perdew}}, \ and\ \bibinfo {author} {\bibfnamefont {G.~I.}\ \bibnamefont
  {Csonka}},\ }\href@noop {} {\bibfield  {journal} {\bibinfo  {journal} {Phys.
  Rev. A}\ }\textbf {\bibinfo {volume} {78}},\ \bibinfo {pages} {022513}
  (\bibinfo {year} {2008}{\natexlab{b}})}\BibitemShut {NoStop}%
\bibitem [{\citenamefont {Quek}\ \emph {et~al.}(2007)\citenamefont {Quek},
  \citenamefont {Venkataraman}, \citenamefont {Choi}, \citenamefont {Louie},
  \citenamefont {Hybertsen},\ and\ \citenamefont {Neaton}}]{QVCL07}%
  \BibitemOpen
  \bibfield  {author} {\bibinfo {author} {\bibfnamefont {S.~Y.}\ \bibnamefont
  {Quek}}, \bibinfo {author} {\bibfnamefont {L.}~\bibnamefont {Venkataraman}},
  \bibinfo {author} {\bibfnamefont {H.~J.}\ \bibnamefont {Choi}}, \bibinfo
  {author} {\bibfnamefont {S.~G.}\ \bibnamefont {Louie}}, \bibinfo {author}
  {\bibfnamefont {M.~S.}\ \bibnamefont {Hybertsen}}, \ and\ \bibinfo {author}
  {\bibfnamefont {J.~B.}\ \bibnamefont {Neaton}},\ }\href@noop {} {\bibfield
  {journal} {\bibinfo  {journal} {Nano Lett.}\ }\textbf {\bibinfo {volume}
  {7}},\ \bibinfo {pages} {3477} (\bibinfo {year} {2007})}\BibitemShut
  {NoStop}%
\bibitem [{\citenamefont {Schmitteckert}\ and\ \citenamefont
  {Evers}(2008)}]{SE08}%
  \BibitemOpen
  \bibfield  {author} {\bibinfo {author} {\bibfnamefont {P.}~\bibnamefont
  {Schmitteckert}}\ and\ \bibinfo {author} {\bibfnamefont {F.}~\bibnamefont
  {Evers}},\ }\href@noop {} {\bibfield  {journal} {\bibinfo  {journal} {Phys.
  Rev. Lett.}\ }\textbf {\bibinfo {volume} {100}},\ \bibinfo {pages} {086401}
  (\bibinfo {year} {2008})}\BibitemShut {NoStop}%
\end{thebibliography}%
\end{document}